\documentclass[9pt,twocolumn,twoside]{pnas-new}
\usepackage{amsmath}
\usepackage{color}
\usepackage[final]{pdfpages}
\newcommand{\new}[1]{{#1}}
\newcommand{\newt}[1]{{{\color{black}#1}}}

\setboolean{displaywatermark}{false}

\templatetype{pnasresearcharticle} 

\title{Landau-Ginzburg theory of cortex dynamics: scale-free
  avalanches emerge at the edge of synchronization}

\author[a,b,c,1]{Serena di Santo} \author[a,1]{Pablo Villegas}
\author[b,c]{Raffaella Burioni} \author[a,2]{Miguel A. Mu\~noz}

\affil[a]{Departamento de Electromagnetismo y F{\'\i}sica de la
  Materia e Instituto Carlos I de F{\'\i}sica Te\'orica y
  Computacional. Universidad de Granada. E-18071, Granada, Spain}
\affil[b]{Dipartimento di Fisica e Scienza della Terra, Universit\`a
  di Parma, via G.P. Usberti, 7/A - 43124, Parma, Italy}
\affil[c]{INFN, Gruppo Collegato di Parma, via G.P. Usberti, 7/A -
  43124, Parma, Italy}

\leadauthor{di Santo} 

\significancestatement{The human cortex operates in a state of
  restless activity, whose meaning and functionality are still not
  well understood.  A fascinating though still controversial
  hypothesis, to some extent backed by empirical evidence, suggests
  that the cortex might work at the edge of a phase transition, from
  which important functional advantages stem. However, the nature of such a
  phase transition is still not fully understood.  Here, we adopt
  ideas from the physics of phase transitions, to construct a general
  (Landau-Ginzburg) theory of cortical networks, allowing us to
  analyze their possible collective phases and phase transitions. We
  conclude that the empirically reported scale-invariant avalanches
  can possibly come about if the cortex operated at the edge of a
  synchronization phase transition, at which neuronal avalanches and
  incipient oscillations coexist.}

\authorcontributions{All authors designed the project, analyzed data,
  and wrote the paper; S. d. S. and P. V. performed the computational
  analyses.}  \authordeclaration{The authors declare no conflict of
  interest.}  \equalauthors{\textsuperscript{1}S.d.S and
  P.V. contributed equally to this work.}
\correspondingauthor{\textsuperscript{2}To whom correspondence should
  be addressed. E-mail: mamunoz@onsager.ugr.es}


\begin{abstract}
  Understanding the origin, nature and functional significance of
  complex patterns of neural activity, as recorded by diverse
  electro-physiological and neuroimaging techniques, is a central
  challenge in Neuroscience.  \new{Such patterns include collective
    oscillations, emerging out of neural synchronization, as well as}
  highly-heterogeneous outbursts of activity interspersed by periods
  of quiescence, called ``neuronal avalanches''.  Much debate has been
  generated about the possible scale-invariance or criticality of such
  avalanches, and its relevance for brain function.  Aimed at shedding
  light onto this, here we analyze the large-scale collective
  properties of the cortex by using a mesoscopic approach, following
  the principle of parsimony of Landau-Ginzburg. Our model is similar
  to that of Wilson-Cowan \new{for neural dynamics} but, crucially,
  including stochasticity and space; synaptic plasticity and
  inhibition are considered as possible regulatory
  mechanisms. Detailed analyses uncover a phase diagram including
  down-states, synchronous, asynchronous, and up-state phases, and
  reveal that empirical findings for neuronal avalanches are
  consistently reproduced by tuning our model to the edge of
  synchronization. This reveals that the putative criticality of
  cortical dynamics does not correspond to a quiescent-to-active phase
  transition, as usually assumed in theoretical approaches, but to a
  synchronization phase transition, at which incipient oscillations
  and scale-free avalanches coexist.  Furthermore, our model also
  accounts for up and down states as they occur, e.g. during deep
  sleep. The present approach constitutes a framework to
  rationalize the possible collective phases and phase transitions of
  cortical networks in simple terms, thus helping shed light into
  basic aspects of brain functioning from a very broad perspective.
  \end{abstract}

\dates{This manuscript was compiled on \today}

\begin{document}


\maketitle
\ifthenelse{\boolean{shortarticle}}{\ifthenelse{\boolean{singlecolumn}}{\abscontentformatted}{\abscontent}}{}

\dropcap{T}he cerebral cortex exhibits spontaneous activity even in
the absence of any task or external stimuli
\cite{Arieli,Raichle,Fox}. A salient aspect of this, so-called,
\emph{resting-state} dynamics, as revealed by \emph{in vivo} and
\emph{in vitro} measurements, is that it exhibits outbursts of
electrochemical activity, characterized by brief episodes of
\new{coherence --during which many neurons fire within a narrow time
  window--} interspaced by periods of relative quiescence, giving rise
to collective oscillatory rhythms \cite{Persi,Segev2}.  Shedding light
on the origin, nature, and functional meaning of such an intricate
dynamics is a fundamental challenge in Neuroscience \cite{Buzsaki}.

Upon experimentally enhancing the spatio-temporal resolution of
activity recordings, Beggs and Plenz made the remarkable finding that,
actually, synchronized outbursts of neural activity could be
decomposed into complex spatio-temporal patterns, thereon named
``neuronal avalanches'' \cite{BeggsPlenz}.  \new{The sizes and
  durations of such avalanches were reported to be distributed as
  power-laws, i.e. to be organized in a scale-free way, limited only
  by network size \cite{BeggsPlenz}. Furthermore, they obey
  finite-size scaling \cite{Peterman2009}, a trademark of scale
  invariance \cite{Binney}, and the corresponding exponents are
  compatible with those of an unbiased branching process
  \cite{Branching}}.

Scale-free avalanches of neuronal activity have been consistently
reported to occur across neural tissues, preparation types,
experimental techniques, scales, and species
\cite{Torre2007,Pasquale2008,Hahn2010,Haimovici,Taglia,Plenz-Shriki2013,2-photon,Linken2012}.
This has been taken as empirical evidence backing the criticality
hypothesis, i.e. the conjecture that the awake brain might extract
essential functional advantages --including maximal sensitivity to
stimuli, large dynamical repertoires, optimal computational
capabilities, etc.-- from operating close to a critical point,
separating two different phases
\cite{Chialvo2004,Schuster,Chialvo2010,Mora-Bialek}.

In order to make further progress, it is of crucial importance to
clarify the nature of the phase transition marked by such an alleged
critical point. \new{It is usually assumed that it corresponds to the
  threshold at which neural activity propagates marginally in the
  network, i.e. to the critical point of a quiescent-to-active phase
  transition \cite{BeggsPlenz}, justifying the emergence of
  branching-process exponents \cite{Henkel,Marro}.  However, several
  experimental investigations found evidence that scale-free
  avalanches emerge in concomitance with collective oscillations,
  suggesting the presence of a synchronization phase transition
  \cite{Gireesh,Yang}.}

\new{From the theoretical side, on the one hand,} very interesting
models accounting for the self-organization of neural networks to the
neighborhood of the critical point of a quiescent-to-active phase
transition have been proposed
\cite{Levina2007,Lucilla2006,Millman,Plenz2015}. These approaches rely
on diverse regulatory mechanisms \cite{homeostatic}, such as synaptic
plasticity \cite{TM96}, spike-time-dependent plasticity
\cite{Shin-Kim}, excitability adaptation, etc. to achieve network
self-organization to \new{the vicinity of} a critical point.
\new{These models} have in common that they rely on \newt{an} extremely large
separation of dynamical timescales (as in models of self-organized
criticality\footnote{This theory, developed three decades ago aims at
  explaining the seemingly ubiquitous presence of criticality in
  natural systems as the result of auto-organization to the critical
  point of a quiescent/active phase transition by means of diverse
  mechanisms, including the presence of two dynamical processes
  occurring at infinitely separated timescales \cite{BJP,JABO1}.}
\cite{Bak,Jensen}) which might not be a realistic assumption
\cite{Levina2007,Lucilla,JABO2,Plenz2015}. \new{Some other models are
  more realistic from a neurophysiological viewpoint}
\cite{Millman,2-photon}, but they give rise to scale-free avalanches
if and only if causal information --which is available in computational
models but not accessible in experiments \cite{Neutral-paper}-- \newt{is considered}. Thus,
in our opinion, a sound theoretical model justifying the empirical
observation of putative criticality is still missing.

On the other hand, from the synchronization viewpoint, well-known
simple models of networks of excitatory and inhibitory spiking neurons
exhibit differentiated synchronous (oscillatory) and asynchronous
phases, with a synchronization phase transition in between
\cite{vanV,Amit-Brunel,Brunel,Brunel-Hakim}. However, avalanches do
not usually appear (or are not searched for) in such modeling
approaches (see, however, \cite{Linken2012,Shew2015,Touboul}).

Concurrently, during deep sleep and also under anesthesia the cortical
state has long been known to exhibit, so called, ``up and down''
transitions between states of high and low neural activity,
respectively \cite{Destexhe2009,Steriade1993}, which clearly deviate
from the possible criticality of the awake brain, and which have been
modeled on their own \cite{Holcman,Mejias,Millman}. \new{Thus, it
  would be highly desirable to design theoretical models describing
  within a common framework the possibility of criticality,
  oscillations, and up-down transitions.}

  Our aim here is to clarify the nature of the phases and phase
  transitions of dynamical network models of the cortex by
  constructing a general unifying theory based on minimal assumptions
  and \new{allowing us, in particular, to elucidate what the nature of
    the alleged criticality is. }

  To construct such a theory we follow the strategy pioneered by
  Landau and Ginzburg.  Landau proposed a simple approach to the
  analysis of phases of matter and the phase transitions they
  experience. It consists in a parsimonious, coarse-grained, and
  deterministic description of states of matter in which --\newt{relying on
  the idea of universality}-- only relevant ingredients (such as
  symmetries and conservation laws) need to be taken into account and
  in which most microscopic details are safely neglected
  \cite{Stanley-book,Binney}.  Ginzburg went a step further by
  realizing that fluctuations are an essential ingredient to be
  included in any sound theory of phase transitions, especially in
  low-dimensional systems. The resulting Landau-Ginzburg theory,
  including fluctuations and spatial dependence \new{is regarded as a
    \emph{meta-model} of phase transitions} \new{and constitutes a}
  firm ground on top of which the standard theory of phases of matter
  rests \cite{Binney}.  Similar coarse-grained theories are nowadays
  used in interdisciplinary contexts -- such as \new{ collective
    motion \cite{Yuhai2005},} population dynamics \cite{Eluding}, and
  neuroscience \cite{Buice,Bressloff1,Bressloff2}-- where diverse
  collective phases stem out of the interactions among many elementary
  constituents.

  In what follows we propose and analyze a Landau-Ginzburg theory for
  cortical neural networks --which can be seen as a variant of the
  well-known Wilson-Cowan model including, crucially, stochasticity
  and spatial dependence-- allowing us to shed light from a very
  general perspective on the collective phases and phase transitions
  that dynamical cortical networks can harbor. Employing analytical
  \new{and, mostly,} computational techniques, we show that our theory
  explains the emergence of scale-free avalanches, as episodes of
  marginal and transient synchronization in the presence of a
  background of ongoing irregular activity, reconciling the
  oscillatory behavior of cortical networks with the presence of
  scale-free avalanches. Last but not least, our approach also allows
  for a \new{unification} of existing models describing diverse
  specific aspects of the cortical dynamics, such as up and down
  states and up-and-down transitions, within a common mathematical
  framework, \new{and is amenable of future theoretical
    (e.g. renormalization group) analyses}.

\section*{Model and Results}
We construct a mesoscopic description of neuronal activity, where the
building blocks are not single neurons but local neural
populations. These latter can be thought as small sections of neural
tissue \cite{Kandel,Dayan} consisting of a few thousand cells (far
away from the large-network limit), and susceptible to be described by
a few variables.  Even though this effective description is
constructed here on phenomenological bases, more formal mathematical
derivations of similar equations from microscopic models exist in the
literature (see e.g. \cite{Benayoun}). In what follows, first (i) we
model the neural activity at a single mesoscopic ``unit'', then (ii)
we analyze its deterministic behavior as a function of parameter
values, and later on (iii) we study the collective dynamics of many
coupled units.

\subsection*{Single-unit model}
At each single unit we consider a dynamical model in which the
excitatory activity, $\rho$, obeys a Wilson-Cowan equation \cite{WC}
(that, following the Landau approach, we truncate to third order in a
Taylor series expansion)\footnote{We keep up to third order to include
  the leading effects of the sigmoid response function; a
  non-truncated variant of the model has also been considered; see SI
  appendix SI1.}:
\begin{equation}
  \dot{\rho}(t)= \big[ -a + R(t)  \big] \rho(t) +  b \rho^2(t) - \rho^{3}(t) + h
\label{rho}
\end{equation}
where $a>0$ controls the spontaneous decay of activity, which is
partially compensated by the generation of additional activity at a
rate proportional to the amount of available \emph{synaptic
  resources}, $R(t)$; \new{the quadratic term with $b>0$, controls
  non-linear integration effects}\footnote{\new{Single neurons
    integrate many presynaptic spikes to go beyond threshold, and thus
    their response is non-linear: the more activity the more likely it
    is self-sustained \cite{Kandel}. As a matter of fact, the
    Wilson-Cowan model includes a sigmoid response function with a
    threshold, implying that activity has to be above some minimum
    value to be self-sustained, and entailing $b>0$ in the series
    expansion (see Methods).}}; finally, the cubic term imposes a
saturation level for the activity, preventing unbounded growth, and
$h$ is an external driving field.

\new{A second equation is employed to describe the dynamics of the
  available synaptic resources, $R(t)$,} through the combined effect of synaptic
depression and synaptic recovery, as encoded in the celebrated model
of Tsodyks and Markram (TM) for synaptic plasticity \cite{TM96,TM97}:
\begin{equation}
\dot{R}(t)=\frac{1}{\tau_R}(\xi-R(t))- \frac{1}{\tau_D} R(t) \rho(t),
\label{R}
\end{equation}
where $\tau_R$ (resp. $\tau_D$) is the characteristic recovery
(depletion) time, and $\xi$ is the baseline level of non-depleted
synaptic resources.  Importantly, we have also considered variants of
this model, avoiding the truncation of the power-series expansion, or
including an inhibitory population as the chief regulatory mechanism:
either of these extensions leads to essentially the same phenomenology
and phases as described in what follows, \new{supporting the
  robustness of the forthcoming results (see Supp. Appendix SI1).}
\begin{figure}
\centering
\includegraphics[width=1.0\columnwidth]{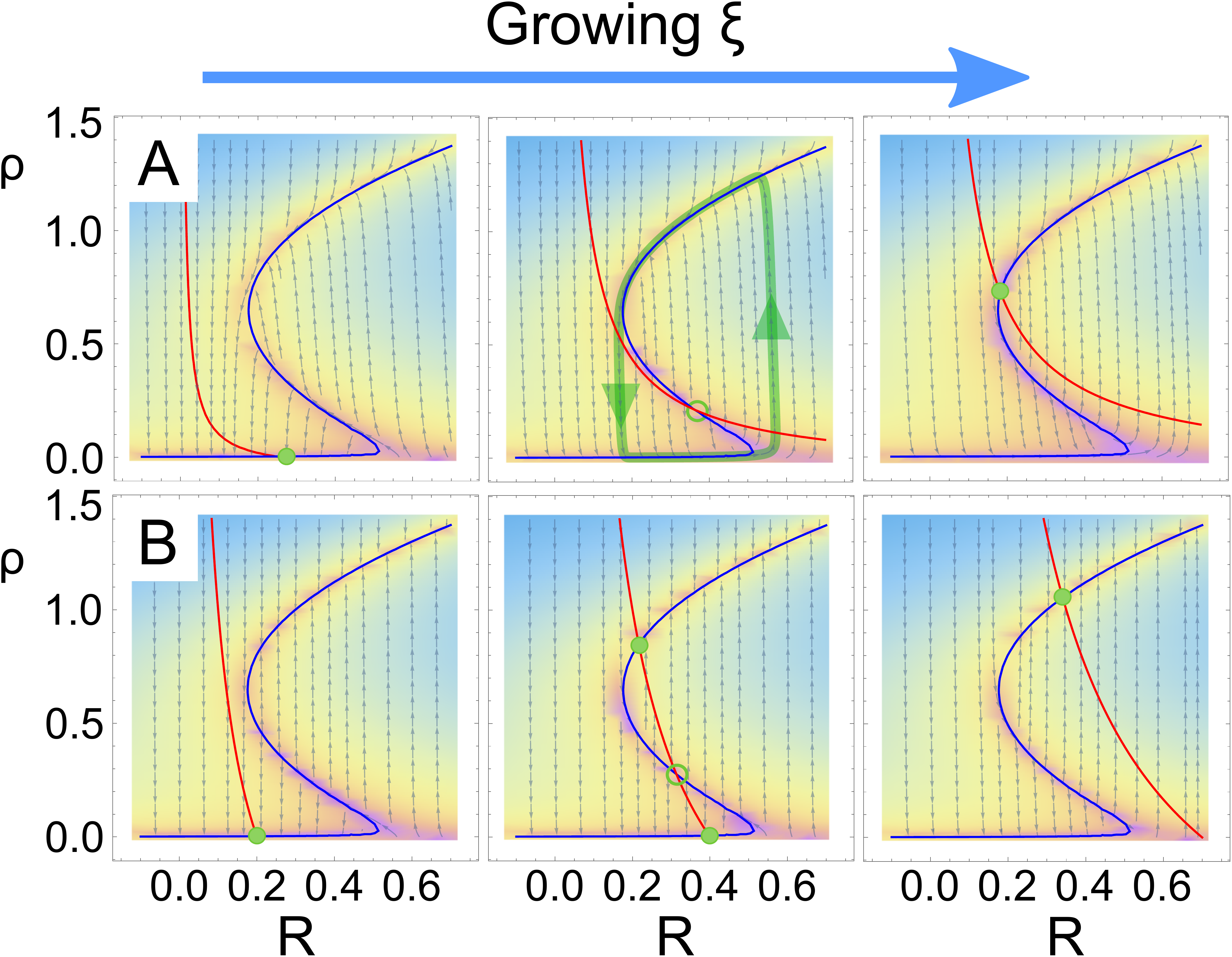}
\caption{{\bf Phase portraits and nullclines for the (deterministic)
    dynamics, Eqs.(\ref{rho}) and (\ref{R}).}  Nullclines are colored
  in blue ($\dot{\rho}=0$) and red ($\dot{R}=0$), respectively; fixed
  points $(\rho^*,R^*)$ --at which nullclines intersect-- are
  highlighted by green full (empty) circles for stable (unstable)
  fixed points. Background color code (shifting from blue to purple)
  represents the intensity of the vector field $(\dot\rho,\dot R)$,
  whose local directions are represented by small grey arrows. A trajectory
  illustrating a limit cycle is showed in green in (A).  The system
  exhibits either (A) an oscillatory regime or (B) a region of
  bistability, in between a down (left) and an up (right) state. It is
  possible to shift from case (A) to case (B) and viceversa by
  changing just one parameter; e.g. the timescale of resources
  depletion, $\tau_D^{-1}$ ($0.016$ and $0.001$ for cases (A) and (B),
  respectively). Other parameter values:
  $h=10^{-3}, a=0.6, b=1.3,\tau_R=10^3$; control parameter, from left
  to right, $\xi=0.3, 1.6, 2.3$ in the upper panel and
  $\xi=0.2, 0.4, 0.7$ in the lower one.}
\label{fig:phase_portrait}
\end{figure}

\subsubsection*{Mean-field analysis}
Here we analyze, both analytically and computationally, the dynamics
of a single unit, as given by Eqs.(\ref{rho}) and (\ref{R}). First, we
determine the fixed points $ (\rho^*,R^*)$ --i.e. the possible
steady-states at which the system can settle-- as a function of the
baseline-level of synaptic resources, $\xi$, which plays the role of a
control parameter (all other parameters are kept fixed to standard
non-specific values, as summarized in the caption of
Fig.\ref{fig:phase_portrait}).  For small values of $\xi$, the system
falls into a quiescent or $down$ state with $\rho^* \approx 0$ and
$R^* \approx \xi$\footnote{Deviations from $\rho^*=0$ stem from the
  small but non-vanishing external driving $h\neq0$.}. Instead, for
large values of $\xi$ there is an active or $up$ state with
self-sustained spontaneous activity $\rho^*>0$ and depleted resources
$R^* <\xi$.  In between these two limiting phases, \new{two
  alternative scenarios (as illustrated in
  Fig.\ref{fig:phase_portrait} and summarized in the phase diagram of
  Suppl. Inf. SI2) can appear depending on the time scales $\tau_D$
  and $\tau_R$:}

(A) A stable limit cycle (corresponding to an unstable fixed point
with complex eigenvalues) emerges for intermediate values of $\xi$ (in
between two Hopf bifurcations) as illustrated in
Fig.\ref{fig:phase_portrait}A.  This Hopf-bifurcation scenario has
been extensively discussed in the literature (see e.g. \cite{Vives})
and it is at the basis of the emergence of oscillations in neural
circuits.

(B) An intermediate regime of bistability including three fixed points
is found for intermediate values of $\xi$ (in between two saddle-node
bifurcations): the up and the down ones, as well as an unstable fixed
point in between (as illustrated in
Fig.\ref{fig:phase_portrait}B). This saddle-node scenario is the
relevant one in models describing transitions between up (active) and
down (quiescent) states \cite{Holcman,Levina2009,Millman}.

Two remarks are in order. The first is that one can shift from one
scenario to the other just by changing one parameter, e.g.  the
synaptic depletion timescale $\tau_D$\footnote{Note that the slope of
  the nullcline deriving from Eq.(\ref{R}) (red in
  Fig.\ref{fig:phase_portrait}) is proportional to $\tau_D$: if it is
  small enough, there exists only one unstable fixed point, giving
  rise to a Hopf bifurcation; otherwise the nullclines intersect at
  three points, generating the bistable regime. These two
  possibilities correspond to cases A and B above, respectively.}. The
second and very important one is that none of these two scenarios
exhibits a continuous transition (transcritical bifurcation)
separating the up/active from the down/quiescent regimes. Thus, at
this single-unit/deterministic level, there is no precursor of a
critical point for marginal propagation of activity.

\subsection*{Stochastic network model}

We now introduce stochastic and spatial effects in the simplest
possible way.  For this, we consider a network of $N$ nodes coupled
following a given connection pattern, as described below. Each network
node represents a mesoscopic region of neural tissue or ``unit'' as
described above. On top of this deterministic dynamics, we consider
that each unit (describing a finite population) is affected by
intrinsic fluctuations
\cite{Benayoun,Bressloff1,Deco-Jirsa}. \new{More specifically,}
Eq.(\ref{rho}) is complemented with an additional term
$+ A(\rho) \eta(t)$ which includes a (zero-mean, unit-variance)
Gaussian noise $\eta(t)$ and a density-dependent amplitude
$A(\rho)$\footnote{In the limit of slow external driving and up to
  leading order in an expansion in powers of $\rho$, this can be
  written as $A(\rho) = \sigma\sqrt{\rho(t)}$, where $\sigma$ is a
  noise amplitude; this stems from the fact that the spiking of each
  single neuron is a stochastic process, and the overall fluctuation
  of the density of a collection of them scales with its square-root,
  as dictated by the central limit theorem \cite{Gardiner} (see also
  \cite{Benayoun} for a detailed derivation of the square-root
  dependence).} i.e. a multiplicative noise \cite{Gardiner}.

\begin{figure*}[h]
\centering
\includegraphics[width=1.0\linewidth]{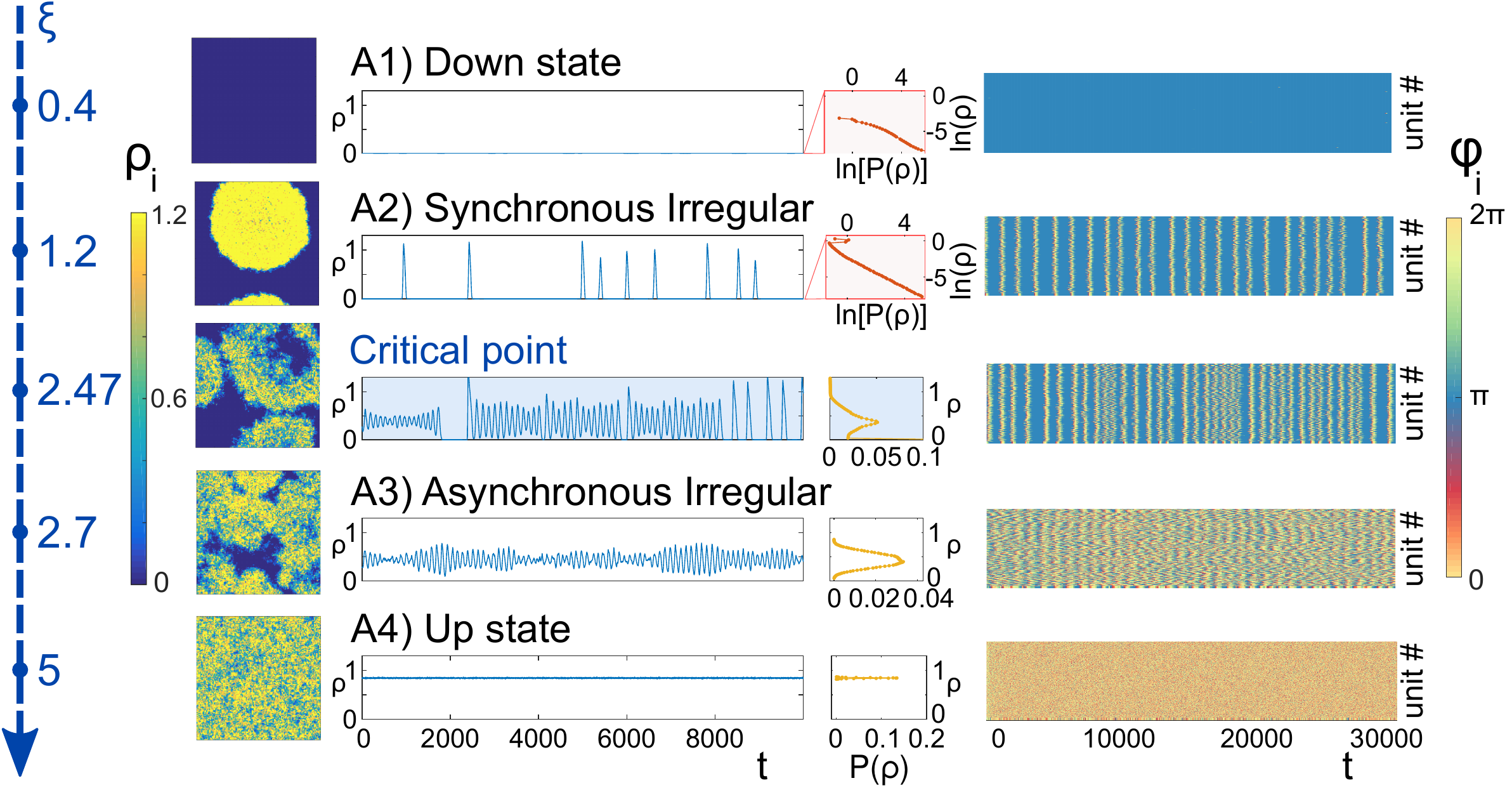}
\caption{ {\bf{Illustration of the diverse phases emerging in the
      model (case A).}}  The baseline of synaptic resources, $\xi$,
  increases from top to bottom: $\xi=0.4 $ (down-state), $\xi=1.2$
  (synchronous regime), $\xi=2.47$ (critical point for the considered
  size, $N=128^2$), $\xi=2.7$ (asynchronous phase), and $\xi=5$
  (active phase).  \emph{First column:} Snapshots of typical
  configurations; the color code represents the level of activity at
  each unit as shown in the scale. The network-spiking or synchronous
  irregular phase, is characterized by waves of activity growing and
  transiently invading the whole system, before extinguishing the
  resources and coming to an end. On the other hand, in the
  nested-oscillation or asynchronous irregular regime multiple
  traveling waves coexist, interfering with each other. In the
  up-state waves are no longer observed and a homogeneous state of
  self-sustained activity is observed (see also the videos in
  SI-Movie).  \emph{Second column:} Time series of the overall
  activity averaged over the whole network.  In the down state
  activity is almost vanishing. In the synchronous phase macroscopic
  activity appears in the form of almost synchronous bursts,
  interspersed by almost silent intervals. At the critical point
  network spikes begin to superimpose, giving rise to complex
  oscillatory patterns (nested oscillations) and marginally
  self-sustained global activity all across the asynchronous regime;
  finally, in the up state the global activity converges to
  steady-state with small fluctuations. \emph{Third column:} Steady
  state probability distribution $P(\rho)$ for the global activity: in
  the down state and the network spiking regime the distributions are
  shown in a double-logarithmic scale; observe the approximate
  power-law for very small values of $\rho$ stemming from the presence
  of multiplicative noise \cite{Branching}. \emph{Fourth column:}
  Illustration of the different levels of synchronization across
  phases: a sample of $200$ randomly chosen units are mapped into
  oscillators using their \emph{analytic-signal representation} (see
  Methods); the plot shows the time evolution of their corresponding
  phases $\phi_k^{\mathcal{A}}$. Observe the almost periodic behavior
  in the synchronous phase, which starts blurring at the critical
  point, and progressively vanishes as the control parameter is
  further increased. Parameter values:
  $a=1, b=1.5, \tau_R=10^3, \tau_D=10^2, h=10^{-7}$.}
\label{Timeseries}
\end{figure*}
At macroscopic scales, the cortex can be treated as a two-dimensional
sheet consisting mostly of short-range connections
\cite{Breakspear-review}\footnote{This type of approach is at the
  bases of, so-called, neural-field models, with a long tradition in
  neuroscience \cite{Deco-review}.}. Although long-range connections
are also known to exist, and small-world effects have been identified
in local cortical regions \cite{Sporns}, here we consider a
two-dimensional square lattice (size $N=L^2$) of mesoscopic units as
the simplest way to embed our model into space.  \newt{Afterward, we
  shall explore how our main results are affected by the introduction
  of more realistic network architectures including additional layers
  of complexity such as long-range connections and spatial
  heterogeneity.}

  Following the parsimonious Landau-Ginzburg approach adopted here,
  the coupling between neighboring units is described up to leading
  order by a diffusion term. This type of diffusive coupling between
  neighboring mesoscopic units stems from electrical synapses
  \cite{Kandel,Torres}, \newt{has some experimental backing \cite{Yu},}
  and has been analytically derived starting from models of spiking
  neurons \cite{Buice} \footnote{More elaborated approaches including
    coupling kernels between different regions, as well as asymmetric
    ones, are also often considered in the literature
    (e.g. \cite{Bressloff2}), but here we stick to the simplest
    possible coupling.}. Thus, finally, the resulting set of coupled
  stochastic equations is:
\begin{equation}
\begin{cases}
  \dot{\rho_i}(t)=(-a + R_i)\rho_i + b\rho_i^2 - \rho_i^{3}+h
  + D \nabla^{2}\rho_i+ \sigma \sqrt{\rho_i}\eta_i\\
  \dot{R_i}(t)=\frac{1}{\tau_R}(\xi-R_i)- \frac{1}{\tau_D} R_i\rho_i
\end{cases}
\label{eq:main}
\end{equation}
where, for simplicity, some time dependences have been omitted;
$\rho_i(t)$ and $R_i(t)$ are, respectively, the activity and resources
at a given node $i$ (with $i=1,2,...N$) and time $t$,
$D \nabla^{2} \rho_i \equiv D \sum_{j\in n.n.i}(\rho_j-\rho_i)$,
describes the diffusive coupling of unit $i$ with its nearest
neighbors $j$, with (diffusion) constant $D$. \new{The physical scales
  of the system are controlled by the values of the parameters $D$ and
  $\sigma$; however, given that, as illustrated in SI2, results do not
  change qualitatively upon varying parameter values (as long as they
  are finite and non-vanishing), here we take $D=\sigma=1$ for the
  sake of simplicity.}

\new{Eq.(\ref{eq:main}) constitutes the basis of our theory. In
  principle, this set of equations is amenable to theoretical
  analyses, possibly including renormalization ones
  \cite{Binney}. However, here we restrict ourselves to computational
  studies aimed at scrutinizing what is the basic phenomenology,
  leaving more formal analyses for the future.}  In particular, we
resort to numerical integration of the stochastic equations
Eq.\ref{eq:main}, which is feasible thanks to the efficient scheme
developed in \cite{Dornic} to deal with multiplicative noise. We
consider $\delta t= 0.01$ as an integration timestep and keep, as
above, all parameters fixed, except for the baseline level of synaptic
resources, $\xi$, which works as a control parameter.

\subsection*{Phases and phase transitions: Case A}
We start analyzing a sets of parameters lying within case A above. We
study the possible phases that emerge as $\xi$ is varied. These are
illustrated in Fig. \ref{Timeseries} where characteristic snapshots,
overall-activity time series, as well as raster plots are
plotted. \new{For a more vivid visualization, we have also generated
  videos of the activity dynamics in the different phases (see
  SI-Movie).}

\subsubsection*{A1) Down-state phase}
If the baseline level $\xi$ is sufficiently small
($\xi \lesssim 0.75$), resources $R$ are always scarce and the system
is unable to produce self-sustained activity (i.e. it is hardly
excitable) giving rise to a down-state phase, characterized by very
small values of the network time-averaged activity
$\bar\rho \equiv \frac{1}{T} \int^{T}_{0} dt \frac{1}{N}\sum_{i=1}^{N}
\rho_i(t)$ for large times $T$ (see Fig.\ref{Timeseries} first
row). The quiescent state is disrupted only locally by the effect of
the driving field $h$, which creates local activity, barely
propagating to neighboring units.

\subsubsection*{A2) Synchronous irregular (SI) phase}
Above a certain value of resource baseline ($\xi \gtrsim 0.75$) there
exists a wide region in parameter space in which activity generated at
a seed point is able to propagate to neighboring units, triggering a
wave of activity which transiently \new{propagates through} the
network until resources are exhausted, activity ceases, and the
recovery process restarts (see Fig. \ref{Timeseries} second row).
Such waves or ``network-spikes'' appear in \new{an oscillatory, though
  not perfectly periodic, fashion}, with an average separation time
that decreases with $\xi$. In the terminology of Brunel \cite{Brunel},
this corresponds to a \emph{synchronous irregular} (SI) state/phase,
since the collective activity is time-dependent (oscillatory) and
single-unit spiking is irregular (as discussed below).  This
wax-and-wane dynamics resembles that of anomalous, e.g. epileptic,
tissues \cite{Hobbs}.

\subsubsection*{A3) Asynchronous irregular (AI) phase}
For even larger values of resource baseline ($\xi \gtrsim 2.15$), the
level of synaptic recovery is sufficiently high \new{as to allow for
  resource-depleted regions to recover before the previous wave has
  come to an end}. Thereby, diverse travelling waves can coexist and
interfere, giving rise to complex collective oscillatory patterns (see
Fig. \ref{Timeseries} fourth row, which is strikingly similar to,
e.g. EEG data of $\alpha-$rhythms \cite{Lopes}). The amplitude of
these oscillations, however, decreases upon increasing network size
(which occurs as many different local waves are averaged and
deviations from the mean tend to be washed away).  This regime can be
assimilated to an \emph{asynchronous irregular} (AI) phase of Brunel
\cite{Brunel} (see below).

\subsubsection*{A4) Up-state phase}
For even larger values of $\xi$, plenty of synaptic resources are
available at all times, giving rise to a state of perpetual activity
with small fluctuations around the mean value
(Fig. \ref{Timeseries} fifth row), i.e. an up state. Let us finally remark,
that as explicitly shown in the SI5, the AI phase and the Up-state
cannot be distinguished in the infinite network-size limit, in which
there are so many waves to be averaged that a homogeneous steady state
emerges on average in both cases.
\begin{figure}[hbtp]
\includegraphics[width=0.9\columnwidth]{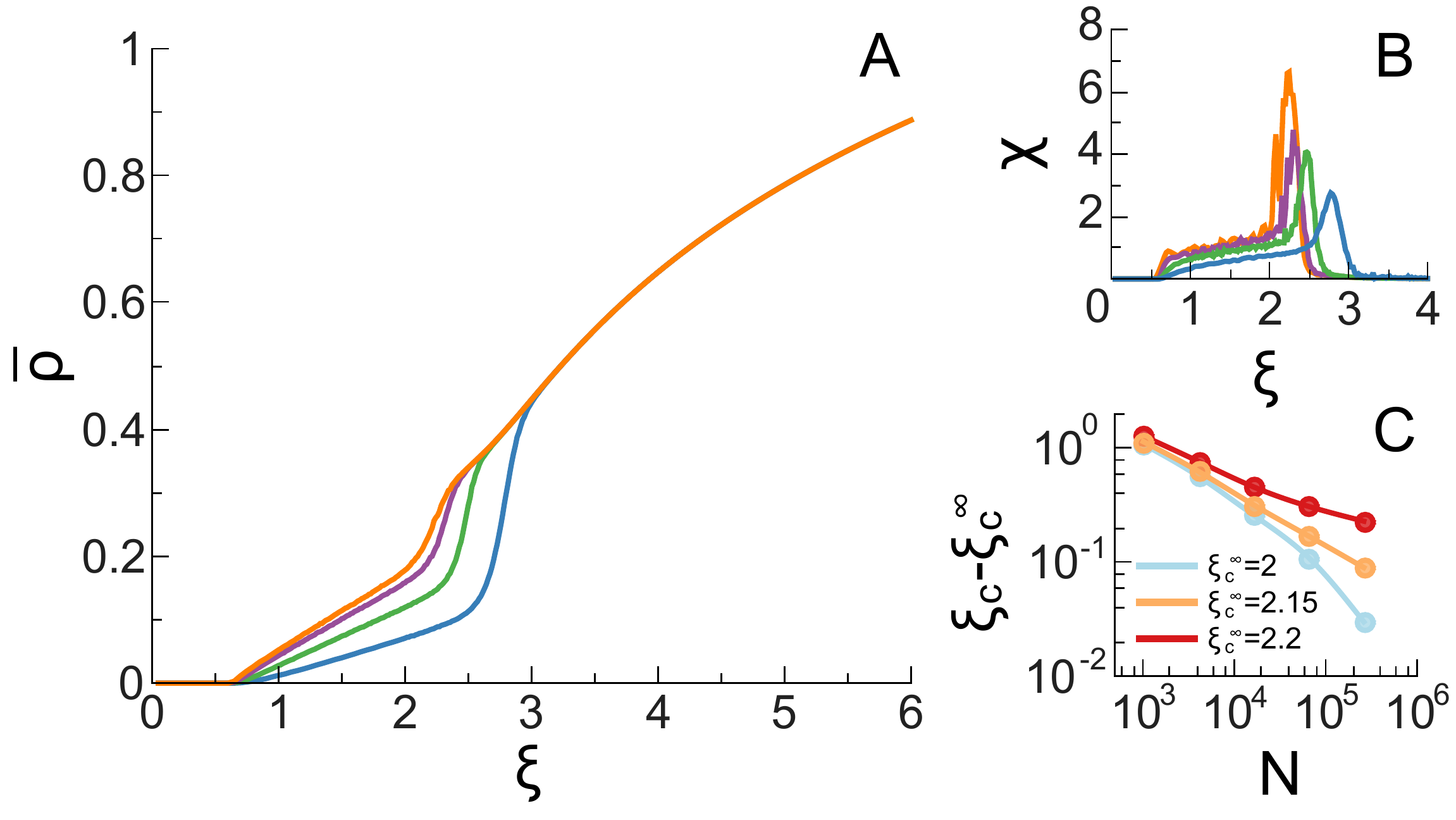}
\caption{{\bf Overall network activity state (case A) as determined by
    the network time-averaged value $\bar\rho$} ($h=10^{-7}$). (A)
  Order parameter $\bar\rho$ as a function of the control parameter
  $\xi$ for various system sizes $N=64^2,128^2,256^2,512^2$ (from
  bottom (blue) to top (orange)); observe that $\bar\rho$ grows
  monotonically with $\xi$ and that an intermediate regime, in which
  $\bar\rho$ grows with system size, emerges between the up and the
  down states.  (B) Standard deviation of the averaged overall
  activity in the system multiplied by $\sqrt{N}$;
  $\Xi=\sigma_{\rho} \sqrt{N}$ (see main text); The point of maximal
  variability coincides with the point of maximal slope in (A) for all
  network sizes $N$.  (C) Finite-size scaling analysis of the peaks in
  (B). The distance of the size-dependent peak locations $\xi_c(N)$
  from their asymptotic value for $N \rightarrow \infty$,
  $\xi_c^{\infty}$, scales as a power law of the system size, taking
  $\xi_c^{\infty} \approx 2.15$, revealing the existence of true
  scaling at criticality.}
\label{fig:phtr}
\end{figure}

\subsubsection*{Phase transitions}
\new{Having analyzed the possible phases, we now discuss the phase
  transitions separating them.}  For all the considered network sizes
the time-averaged overall activity, $\bar{\rho}$, starts taking a
distinctively non-zero value above $\xi \approx 0.75$ (see
Fig.\ref{fig:phtr}), reflecting the upper bound of the down or
quiescent state (transition between A1 and A2).  This phase transition
is rather trivial and corresponds to the onset on network spikes
i.e. oscillations \newt{(whose characteristic time depends on various
  factors, such as the synaptic recovery time \cite{Tabak2000} and the
  baseline level of synaptic resources)}.

More interestingly, Fig.\ref{fig:phtr} also reveals that $\bar{\rho}$
exhibits an abrupt increase at \new{ (size-dependent)} values of
$\xi$, between $2$ and $3$, signaling the transition from A2 to
A3. However, the jump amplitude decreases as $N$ increases, suggesting
a smoother transition in the large-$N$ limit. Thus it is not clear
\emph{a priori}, using $\bar{\rho}$ as an order parameter, whether
there is a true sharp phase transition or there is just a crossover
between the synchronous (A2) and the asynchronous (A3) regimes. To
elucidate the existence of a true critical point, we measured the
standard deviation of the network-averaged global activity
$\bar{\rho}$, $\sigma_\rho$.  \new{Direct application of the central
  limit theorem \cite{Gardiner} would imply that such a quantity
  should decrease as $1/\sqrt{N}$ for large $N$ and thus,
  $\chi \equiv\sqrt{N}\sigma_\rho$ should converge to a constant.
  However, Fig. \ref{fig:phtr}B shows} that $\chi$ exhibits a very
pronounced peak located at the ($N$-dependent) transition point
between the A2 and the A3 phases; furthermore its height grows with
$N$ --i.e. it diverges in the thermodynamic limit-- revealing strong
correlations and anomalous scaling, as occurs at critical points.
Also, a finite-size scaling analysis of the value of $\xi$ at the peak
(for each $N$), i.e. $\xi_c(N)$, reveals the existence of finite-size
scaling, as corresponds to a \emph{bona fide} continuous phase
transition at $\xi_{c}^{\infty}\simeq2.15(5)$ in the infinite-size
limit (see Fig. \ref{fig:phtr}C). Moreover, a detrended
fluctuation analysis \cite{Peng1994,Linkenkaer2001} of the timeseries
reveals the emergence of long-range temporal correlations right at
$\xi_c$ (see SI4), as expected at a continuous phase transition.

\begin{figure}[hbtp]
  \includegraphics[width=0.9\columnwidth]{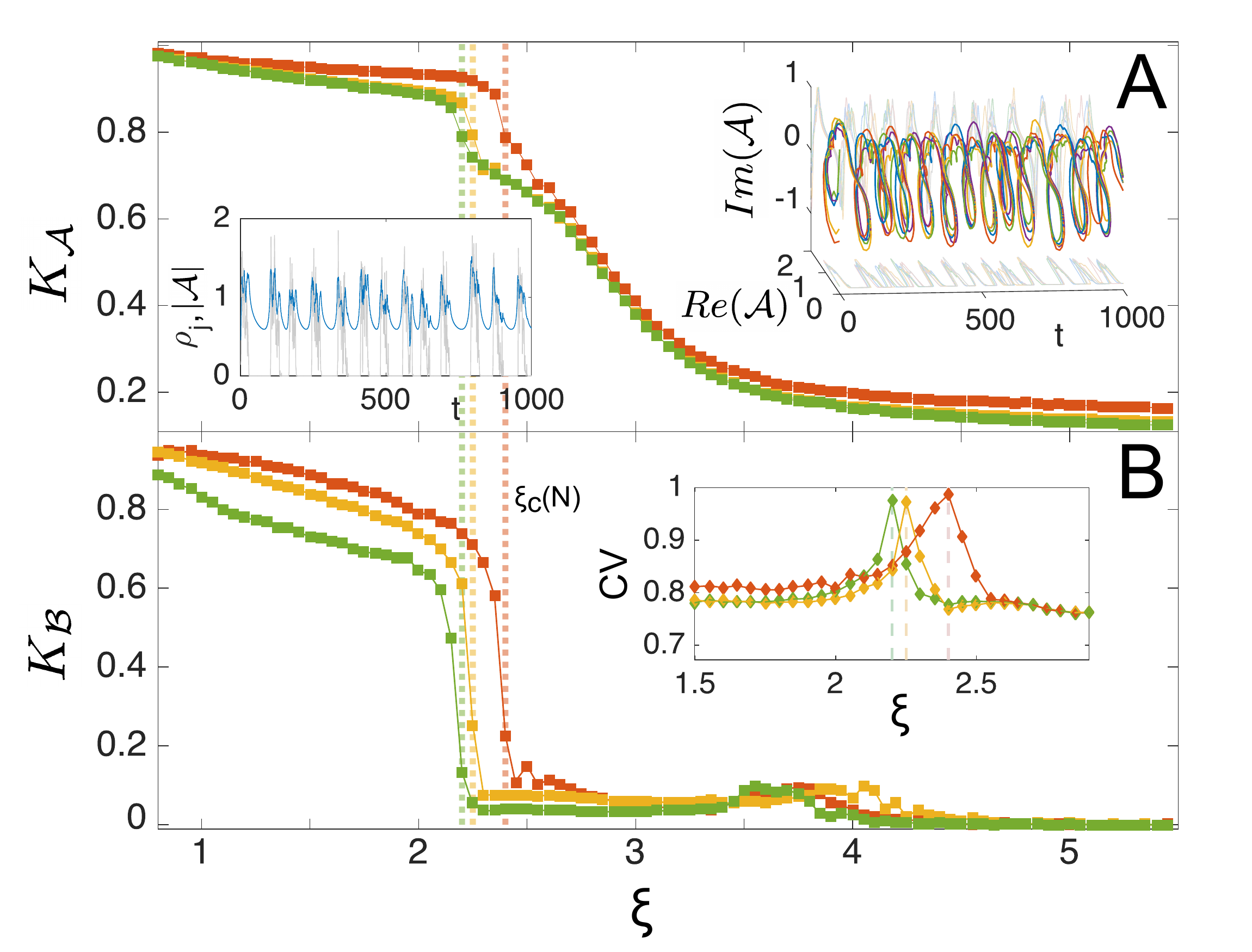}
  \caption{{\bf Synchronization transition elucidated by measuring the
      Kuramoto parameter} as estimated using (A) the analytic signal
    representation ${\mathcal{A}_k(t)}$ of activity time series
    $\rho_k(t)$ at different units $k$ and for various system sizes
    ($N=128^2$ (red), $256^2$ (orange), $512^2$ (green)). For
    illustrative purposes, the top right inset of (A) shows the
    analytical representation (including both a real and an imaginary
    part) of $5$ sample units as a function of time; the inset on the
    left shows the time evolution of one node (gray) together with the
    amplitude of its analytic representation (blue)
    Both insets, vividly illustrate the oscillatory nature of the unit
    dynamics.  (B) \new{Results similar to those of} (A), but
    employing a different method to compute time-dependent phases of
    effective oscillators (see Methods). This alternative method
    captures more clearly the emergence of a transition; the point of
    maximum slope of the curves corresponds to the value of the
    transition points $\xi_c(N)$ in (A).  The inset in (B) shows the
    coefficient of variation $CV$ (ratio of the standard deviation to
    the mean) of the times between two consecutive crossings of the
    value $2 \pi$; it exhibits a peak of variability at the critical
    point $\xi_c(N)$. }
\label{Synchro}
\end{figure}
To shed further light on the nature of such a transition, it is
convenient to employ a more adequate (synchronization)
order-parameter. In particular, we consider the Kuramoto index $K$
--\new{customarily employed to detect synchronization transitions}
\cite{Pikovsky-book}-- defined as
$ K \equiv \frac{1}{N} \left \langle \left| \sum_{k=1}^{N} e^{i\phi_k
      (t)} \right| \rangle \right. $ --where $i$ is the imaginary
unit, $\lvert\cdot \rvert $ is the modulus of a complex number,
$\langle \cdot \rangle$ here indicates averages over time and
independent realizations, and $k$ runs over units, each of which is
characterized by a phase, $\phi_k (t) \in[0,2\pi]$, that can be
defined in different ways. For instance, an effective phase
$\phi_k^{\mathcal{A}}(t)$ can be assigned to the time-series at unit
$k$, $\rho_{k}(t)$, by computing its \emph{analytic signal
  representation}, which maps any given real-valued timeseries into an
oscillator with time-dependent phase and amplitude (see Methods).
Using the resulting phases, $\phi_k^{\mathcal{A}}(t)$, the Kuramoto
index $K_{\mathcal{A}}$ can be calculated.  As illustrated in
Fig. \ref{Synchro}A, it reveals the presence of a synchronization
transition: the value of $K_{\mathcal{A}}$ clearly drops, at the
previously determined critical point $\xi_c(N)$.  An alternative
method to define a time-dependent phase for each unit (details
discussed in Methods) reveals even more vividly the existence of a
synchronization transition at $\xi_c(N)$ as shown in
Fig. \ref{Synchro}B.  Finally, we have also estimated the coefficient
of variation (CV) of the distance between the times at which each of
these effective phases crosses the value $2\pi$; this analysis reveals the
presence of a sharp peak of variability, converging for large network
sizes to the critical point $\xi_c^{\infty}\approx 2.15$ (see inset of
Fig. \ref{Synchro}B).  \vspace{0.5cm}

\begin{figure*}[h]
\centering
\includegraphics[width=0.8\linewidth]{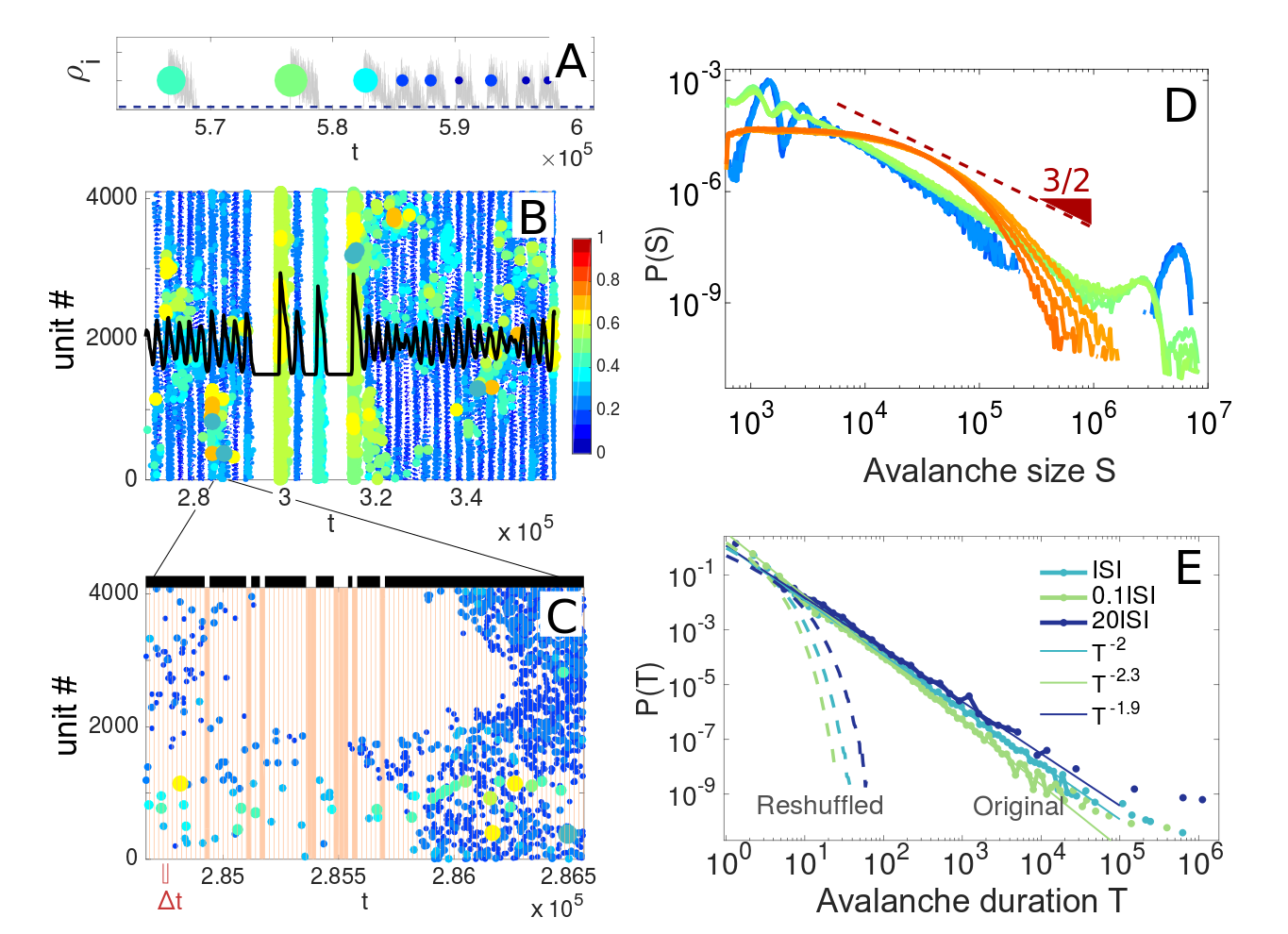}
\caption{{\bf Avalanches measured from activity time series.}  (A)
  Illustration of the activity timeseries $\rho_i(t)$ (grey color) at
  a given unit $i$. By establishing a threshold value $\theta$ (dashed
  blue line, close to the origin) a single ``event'' or ``unit spike''
  is defined at the time of the maximal activity in between two
  threshold crossings (n.b. the forthcoming results are robust to
  changes in this criterion; see SI6); a weight equal to the area
  covered in between the two crossings is assigned to each event (note
  the color code). This allows us to map a continuous time-series into
  a discrete series of weighted events.  The time distance between two
  consecutive events is called \emph{inter-spike interval} ($ISI$).
  (B) Raster plot for a system with $64^2$ units, obtained using the
  procedure above for each unit. Observe that large events coexist
  with smaller ones, and that these last ones, occur in a rather
  synchronous fashion. The overall time-dependent activity is marked
  with a black curve. (C) Zoom of a part of (B) illustrating the time
  resolved structure and using a time binning $\Delta t$ equal to the
  network-averaged $ISI$. Shaded columns correspond to empty time
  bins, i.e. with no spike. Avalanches are defined as sequences of
  events occurring in between two consecutive empty time bins and are
  represented by the black bars above the plot. (D) Avalanche-size
  distribution (the size of the avalanche is the sum of the weighted
  spikes it comprises) for diverse values of $\xi$ (from $1.85$ to
  $2.05$, in blueish colors, from $2.7$ to $2.9$ in greenish colors,
  and from $3.3$ to $3.45$ in orangish colors) measured from the
  raster plot $\Delta t= ISI$.  The (red) triangle, with slope $3/2$
  is plotted as a reference, illustrating that, near criticality, a
  power law with an exponent similar to the experimentally measured
  one is recovered.  Away from the critical point, either in the
  synchronous phase (blueish colors) and the asynchronous one
  (orangish) clear deviations from power-law behavior are
  observed. Observe the presence of ``heaps'' in the tails of the
  distributions, especially in the synchronous regime; these
  correspond to periodic waves of synchronized activity (see SI7);
  they also appear at criticality, but at progressively larger values
  for larger system sizes.  (E) Avalanche-duration distribution,
  determined with different choices of the time bin. The
  experimentally measured exponent $\approx2$ is reproduced using
  $\Delta t=ISI$, whereas deviations from such a value are measured
  for smaller (larger) time bins, in agreement with experimentally
  reported results. After reshuffling times, the distributions become
  exponential ones, with characteristic timescales depending on
  $\Delta t$ (dashed lines).}
\label{Raster-avalanches}
\end{figure*}
Thus, recapitulating, the phase transition separating the down state
from the synchronous irregular regime (A1-A2 transition) is trivial
and corresponds to the onset of network spikes, with no sign of
critical features. In between the asynchronous and the up state
(A3-A4) there is no true phase transition, as both phases are
indistinguishable in the infinitely-large-size limit (see
SI5). \new{On the other hand, different measurements clearly reveal
  the existence of a bona fide synchronization phase transition
  (A2-A3) at which non-trivial features characteristic of criticality
  emerge.}

\subsubsection*{Avalanches}
For ease of comparison with empirical results, we define a protocol to
analyze avalanches \new{ that closely resembles} the experimental one,
as introduced by Beggs and Plenz \cite{BeggsPlenz}.  Each activity
timeseries of an individual unit can be mapped into a series of
discrete-time ``spikes'' or ``events'' as follows. As illustrated in
Fig.\ref{Raster-avalanches}A, a ``spike'' corresponds to a period in
which \new{the activity at a given unit} is above a given small
threshold in between two windows of quiescence (activity below
threshold).\footnote{Results are quite robust to the specific way in
  which this procedure is implemented. See the Methods section as well
  as the caption of Fig.\ref{Raster-avalanches} and Supp. Inf.  SI6).}
Hence, as illustrated in Fig.\ref{Raster-avalanches}B, the network
activity can be represented as a raster plot of spiking units.
Following the standard experimental protocol a discrete time binning
$\Delta t$ is chosen and each individual spike is assigned to one such
bin. \new{An avalanche is defined as a consecutive sequence of
  temporally-contiguous occupied bins} preceded and ended by empty
bins (see Fig.\ref{Raster-avalanches}B and C). Quite remarkably,
using this protocol several well-known experimental key features
\new{of neuronal avalanches can be faithfully reproduced by tuning
  $\xi$} to a value close to the synchronization transition. In
particular:

(i) The sizes and durations of avalanches of activity are found to be
broadly (power-law) distributed at the critical point; these
scale-invariant avalanches coexist with anomalously large events or
``waves'' of synchronization, as revealed by the ``heaps'' in the
tails of the curves of in Fig.\ref{Raster-avalanches}D and E.

(ii) Changing $\Delta t$, power-law distributions with varying
exponents are obtained at criticality (the larger the time bin, the
smaller the exponent) as originally observed experimentally by Beggs
and Plenz (Fig.\ref{Raster-avalanches}E).

(iii) \new{In particular,} when $\Delta t$ is chosen to be equal to
the $ISI$ (inter-spike time interval, i.e. the time interval between
any two consecutive spikes), avalanche sizes and durations obey --at
criticality-- finite-size scaling with exponent values compatible with
the standard ones, i.e. those of an unbiased branching process (see
Fig.\ref{Raster-avalanches}B and C as well as Supporting Information
SI6).

(iv) Reshuffling the times of occurrence of unit's spikes, the
statistics of avalanches is dramatically changed, giving rise to
exponential distributions (as expected for an uncorrelated Poisson
point process) thus revealing the existence of a non-trivial temporal
organization in the dynamics (Fig.\ref{Raster-avalanches}E).

(v) Away from the critical point, both in the sub-critical and in the
supercritical regime, deviations from this behavior are observed; in
the subcritical or synchronous regime, the peak of periodic large
avalanches becomes much more pronounced, while in the asynchronous
phase, such a peak is lost and distribution functions become
exponential ones with a characteristic scale (see
Fig.\ref{Raster-avalanches}D).

Summing up, \new{our model tuned to the edge of a
  synchronization/desynchronization phase transition reproduces all
  chief empirical findings for neural avalanches.} These findings
strongly \new{suggest} that the critical point alluded by the
criticality hypothesis of cortical dynamics does not correspond to a
quiescent/active phase transition --as modeling approaches usually
assume-- but to a synchronization phase transition, at the edge of
which oscillations and avalanches coexist.

\new{It is important to underline that our \newt{results regarding the
    emergence of scale-free avalanches} are purely computational. To
  date, we do not have a theoretical understanding of why results are
  compatible with branching-process exponents. In particular, it is
  not clear to us if a branching process could possibly emerge as an
  effective description of the actual (synchronization) dynamics in
  the vicinity of the phase transition, or whether the exponent values
  appear as a generic consequence of the way temporally-defined
  avalanches are measured (see \cite{Touboul}). These issues deserve
  to be carefully scrutinized in future work.}

\subsubsection*{The role of heterogeneity}

\new{Thus far we have described homogeneous networks with local
  coupling. However, long-range connections among local regions also
  exist in the cortex, and mesoscopic units are not necessarily
  homogeneous across space \cite{Topology,Sporns}. These empirical
  facts motivated us to perform additional analysis of our theory, in
  which slightly modified substrates are employed. First, we
  considered small-world networks, and verified that our main results
  (i.e. the existing phases and phase transitions) are insensitive to
  the introduction of a small percentage of long-range connections
  (see SI3). However, details such as the boundaries of the phase
  diagram, the shape of propagation waves, and the amplitude of nested
  oscillations do change.}

\new{More remarkably, as described in detail in SI3, a simple
  extension of our theory in which parameters are not taken to be
  homogeneous but position-dependent, i.e. heterogeneous in space, is
  able to reproduce remarkably well empirical \emph{in vitro} results
  for neural cultures with different levels of mesoscopic structural
  heterogeneity \cite{Okujeni}.}

\newt{To further explore the influence of network architecture onto
dynamical phases, in future work we will extend our model employing
empirically-obtained large-scale networks of the human brain, as their
heterogeneous and hierarchical-modular architecture is known to
influence dynamical process operating on them \cite{Sporns,Moretti}.}

\subsection*{Phases and phase transitions: Case B}
Here, we discuss the much simpler scenario for which the
deterministic/mean-field dynamics predicts bistability, i.e. case B
above, which is obtained e.g. considering a slower dynamic for
synaptic-resource depletion. In this case, the introduction of noise
and space, does not significantly alter the deterministic
picture. Indeed, computational analyses reveal that there are only two
phases: a down state and an up one for small and large values of
$\xi$, respectively. These two phases have the very same features as
their corresponding counterparts in case A. \new{However,} the phase
transition between them is discontinuous (much as in Fig. 1B) and
thus, for finite networks, fluctuations induce spontaneous transitions
between the up and the down state when $\xi$ takes intermediate
values, in the regime of phase coexistence.  Thus, in case B, our
theory constitutes a sound Landau-Ginzburg description of existing
models, such as those in \cite{Millman,Holcman,Mejias}, describing up
and down states \new{and up-and-down transitions.}

\section*{Conclusions and Discussion}
The brain of mammalians is in a state of \newt{perennial activity even
  in the absence of any apparent stimuli or task}. Understanding the
origin, meaning, and functional significance of such an energetically
costly dynamical state are fundamental problems in neuroscience.  The
--so called-- criticality hypothesis conjectures that the underlying
dynamics of cortical networks is such that it is posed at the edge of
a continuous phase transition, separating qualitatively different
phases or regimes, with different degrees of order. Experience from
statistical physics and the theory of phase transitions teaches that
critical points are rather singular locations in phase diagrams, with
very remarkable and peculiar features, such as scale invariance,
i.e. the fact that fluctuations of wildly diverse spatio-temporal
scales can emerge spontaneously, allowing the system dynamics to
generate complex patterns of activity in a simple and natural way. A
number of features of criticality, including scale invariance, have
been conjectured to be functionally convenient and susceptible to be
exploited by biological (as well as artificial) computing devices.
Thus, the hypothesis that the brain actually works at the borderline
of a phase transition has gained momentum in recent years
\cite{Schuster,Chialvo2010,Mora-Bialek}, even if some skepticism
remains \cite{Touboul}.  However, what these phases are, and what the
nature of the putative critical point is, are questions that still
remain to be fully settled.

Aimed at shedding light on these issues, here we followed a classical
statistical-physics approach. Following the parsimony principle of
Landau and Ginzburg in the study of phases of the matter and the phase
transitions they experience, we proposed a simple stochastic
mesoscopic theory of cortical dynamics that allowed us to classify the
possible emerging phases of cortical networks under very general
conditions. For the sake of specificity and concreteness we focused on
a regulatory dynamics --preventing the level of activity to explode--
controlled by synaptic plasticity (depletion and recovery of synaptic
resources), but analogous results \new{have been obtained} considering
e.g. inhibition as the chief regulatory mechanism. As a matter of
fact, our main conclusions are quite robust and general and do not
essentially depend on specific details of the implementation, the
nature of the regulatory mechanism.

The mesoscopic approach upon which our theory rests is certainly not
radically novel as quite a few related models exist in the
literature. For instance, neural-mass
\cite{Freeman,Friston,Destexhe} and neural-field models
\cite{Deco-review,Breakspear-review}, rate or population
activity equations \cite{Dayan,Gerstner}, are similar in spirit, and
have been successfully employed to analyze activity of populations of
neurons and synapses, and their emerging collective regimes, \new{at
  mesoscopic and macroscopic scales.}

Taking advantage of experience from the theory of phase transitions,
we introduce two important key ingredients: intrinsic stochasticity
stemming from the non-infinite size of mesoscopic regions, and spatial
dependence. In this way, our theory consists of a set of stochastic
\new{(truncated)} Wilson-Cowan equations and can be formulated as a
field theory, employing standard techniques \cite{Tauber}. A rather
similar (field theoretic) approach to analyze fluctuation effects in
extended neural networks \newt{has been proposed} \cite{Buice}.

Such a theory turns out to include a continuous phase transition from
a quiescent to an active phase, with a critical point in between,
which is in contrast with our findings here.  Note, however, that
the authors of \cite{Buice} themselves open the door to more complex
scenarios if refractoriness and thresholds are included.

In any case, such a continuous-phase-transition picture can be easily
recovered in our framework, just by changing the sign of a parameter:
i.e. taking $b<0$ in Eq.(\ref{rho}); with such a parameter choice, our
theory constitutes a sound Landau-Ginzburg description of microscopic
models of neural dynamics exhibiting criticality and a continuous
phase transition from a quiescent to an active phase
\cite{Levina2007,JABO2}. We believe, however, that this scenario does
not properly capture the essence of cortical dynamics as, in actual
networks of spiking neurons, there are spike-integration mechanisms,
meaning that many inputs are required to trigger further activity.

Using our Landau-Ginzburg approach, we have shown that the stochastic
and spatially extended neural networks can harbor two different
scenarios depending on parameter values: case (A) including a limit
cycle at the deterministic level and the possibility of oscillations
and case (B) leading to bistability (see Fig.1).

In the simpler case (B) our complete theory generates either a down or a
homogeneous up-state phase, with a discontinuous transition separating
them, and the possibility of up-down transitions when the system
operates in the bistability region.  In this case, our theory
constitutes a sound mesoscopic description of existing microscopic
models for up-and-down transitions
\cite{Millman,Holcman,Levina2009,Pittorino}.

On the other hand, in case (A), we find diverse phases including
oscillatory and bursting phenomena: down states, synchronous
irregular, asynchronous irregular, and active states.

As a side remark, let us emphasize that we constructed a
coarse-grained model for activity propagation, but our analyses
readily revealed the emergence of oscillations and synchronization
phenomena.  Hence, our results justify the use of models of effective
coupled oscillators to scrutinize the large-scale dynamics of brain
networks. As a matter of fact, such models \new{have been reported} to
achieve the best performance --e.g.  reproducing empirically-observed
resting-state networks \cite{Sporns}-- when operating close to the
synchronization phase transition point
\cite{Cabral,Deco-Jirsa,Villegas1}.

Within our framework, it is possible to define a protocol to analyze
avalanches, resembling very closely the experimental one
\cite{BeggsPlenz,Torre2007,Peterman2009,Plenz-Shriki2013,2-photon}.
Thus, in contrast with other computational models, causal information
is not explicitly needed/employed here to determine avalanches --they
are determined from raw data-- and results can be
straightforwardly compared to experimental ones for neuronal
avalanches, without conceptual gaps \cite{Neutral-paper}.

The model reproduces all the main features observed experimentally:
(i) \new{Avalanche sizes and durations distributed in a scale-free way
  emerge at the critical point of the synchronization transition.}
(ii) The corresponding exponent values depend on the time bin
$\Delta t$ required to define avalanches, but (iii) fixing $\Delta t$
to coincide with the inter-spike interval, \new{$ISI$}, the same
statistics as in empirical networks, i.e. the critical exponents
compatible with those of an unbiased branching process
\cite{Branching} are obtained. Finally (iv) scale-free distributions
disappear if events are reshuffled in time, revealing a non-trivial
temporal organization.

Thus, the main outcome of our analyses is that the underlying phase
transition at which scale-free avalanches emerge does not separate a
quiescent state from a fully active one; \newt{instead, it is} a
synchronization phase transition. This is a crucial observation, as
most of the existing modeling approaches for critical avalanches in
neural dynamics to date rely on a continuous quiescent/active phase
transition.

Consistently with our findings-- the amazingly detailed model put
together by the Human Brain Project consortium suggests that the model
best reproduces experimental features when tuned near to its
synchronization critical point \cite{HBP}. In such a study, the
concentration of Calcium ions, $Ca^{2+}$ needs to be carefully tuned
to its actual nominal value to set the network state. Similarly, in
our approach, the role of the calcium concentration is played by the
parameter $\xi$, regulating the maximum level reachable by synaptic
resources. Interestingly, the calcium concentration is well-known to
modulate the level of available synaptic resources
(i.e. neurotransmitter release from neurons; see
e.g. \cite{TM96,TM97,Dayan}), hence, both quantities play a similar
role.

Observe that here, we have not made any attempt to explore how could
potentially the network self-organize to operate in the vicinity of
the synchronization critical point \new{without the need of parameter
  tuning.} Adaptive, homeostatic and self-regulatory mechanisms
accounting for this will be analyzed in future work. \new{Also, here
  we have not looked for the recently uncovered \emph{neutral neural
    avalanches} \cite{Neutral-paper}, as these require causality
  information to be considered, and such detailed causal relationships
  are blurred away in mesoscopic coarse grained descriptions.}

Summing up, our Landau-Ginzburg theory with parameters lying in case
(B) constitutes a sound description of the cortex during deep sleep or
during anesthesia, when up and down transitions are observed. On the
other hand, case (A) when tuned close to the synchronization phase
transition can be a sound theory for the awaked cortex, in a state of
alertness. A detailed analysis of how the transition between
deep-sleep (described by case (B)) and awake (or REM sleep, described
by case (A)) \newt{states} may actually occur in these general terms is beyond our
scope here, but observe that, just by modifying the speed at which
synaptic resources recover it is possible to shift between the two
cases, making it possible to investigate how such transitions could be
induced.

A simple extension of our theory, including spatial heterogeneity has
been shown to be able to reproduce remarkably well experimental
measurements of activity in neural cultures with structural
heterogeneity, opening the way to more stringent empirical validations
of the general theory proposed here.

Even if further experimental, computational and analytical studies
would be certainly required to definitely settle the controversy about
the possible existence, origin, and functional meaning of the possible
phases and phase transitions in cortical networks, we hope that the
general framework introduced here --based on very general and robust
principles-- helps in clarifying the picture and in paving the way to
future developments in this fascinating field.

\matmethods{
\small
\subsection*{Model details} \footnotesize{In the Wilson-Cowan model,
  in its simplest form, the dynamics of the average firing rate or
  global activity, $\rho$, is governed by the equation
  $$\dot{\rho}(t)=-a \rho\left(t\right)+ 
  (1-\rho(t)) S\left(W\rho\left(t\right)-\Theta\right)$$ where $W$ is the synaptic
  strength, $\Theta$ is a threshold value that can be fixed to unity,
  and $S(x)$ is a sigmoid (transduction) function, e.g.
$S(x)=\tanh(x)$ 
\cite{Benayoun,WC}. We adopt this well-established model and, for
simplicity, keep only the leading terms in a power-series expansion,
and rename the constants, yielding the deterministic part of
Eq.(\ref{rho}).  To this we add noise $\sqrt{\rho(t)}\eta(t)$ --which
is a delta-correlated Gaussian white noise of zero mean and unit
variance, accounting for stochastic/demographic effects in finite
local populations as dictated by the central limit theorem; a formal
derivation of such an intrinsic or demographic noise, starting from a
discrete microscopic model can be found in \cite{Benayoun}). A noise
term could be also added to the equation for synaptic resources
\cite{Mejias}, but it does not significantly affect the results.
Considering $N$ mesoscopic units, and coupling them diffusively within
some networked structure (e.g. a two dimensional lattice), we finally
obtained the set of Eqs.(\ref{eq:main}).}  \small
\subsection*{Analytic signal representation}
\footnotesize{The Hilbert transform $\mathcal{H}(\cdot)$ is a bounded
  linear operator largely used in signal analysis as it provides a
  tool to transform a given real-valued function $u(t)$ into a complex analytic
  function, called  the \emph{analytic signal representation}. This is
  defined as $\mathcal{A}_{u}(t)=u(t)+i\mathcal{H}[u(t)]$ where the
  Hilbert transform of $u(t)$ is given by:
  $ \mathcal{H}[u(t)]=h*u=\frac{1}{\pi}\lim_{\epsilon\rightarrow0}
  \int_{\epsilon}^{\infty}\frac{u(t+\tau)-u(t-\tau)}{\tau}d\tau$.
  Expressing the analytic signal in terms of its time-dependent
  amplitude and phase (polar coordinates) makes it possible to
  represent any signal as an oscillator.  In particular, the associated
  phase is defined by
  $\phi_k^{\mathcal{A}}=\arctan{Im(\mathcal{A}_k)/Re(\mathcal{A}_k)}$.
} \small
\subsection*{From continuous timeseries to discrete events}
\footnotesize{Local timeseries at each single unit, $\rho_k(t)$, can
  be mapped into time sequences of point-like (``unit spiking'')
  events.  For this, a local threshold $\theta\ll1$ is defined,
  allowing to assign a state on/off to each single unit/node
  (depending on whether it is above/below such a threshold) at any
  given time. If the threshold is low enough, the procedure is
  independent of its specific choice. A single (discrete) ``event'' or
  ``spike'' can be assigned to each node $i$, e.g. at the time of the
  maximal $\rho_i$ within the on-state; a weight proportional to the
  integral of the activity time series spanned between two consecutive
  threshold crossings is assigned to each single event (see
  Fig.\ref{Raster-avalanches}A).  Other conventions to define an event
  are possible, but results are not sensitive to it as illustrated in
  the SI6. }  \small
    \subsection*{Phases from spiking patterns}
    \footnotesize{An alternative method to define a phase at each unit
      can be constructed after a continuous timeseries has been mapped
      into a spiking series. In particular, \new{using a linear
        interpolation:}
      $\phi_k^{(\mathcal{B})} (t)=2\pi (t-t^k_n)/(t^k_{n+1}-t^k_n)$
      where $t\in[t^k_n,t^k_{n+1})$ and $t^k_n$ is the time of the
      $n^{th}$ spike of node/unit $k$.}  } \showmatmethods

\acknow{We acknowledge the Spanish-MINECO grant FIS2013-43201-P (FEDER
  funds) for financial support. We are very thankful to P. Moretti,
  J. Hidalgo, J. Mejias, J.J. Torres, A. Vezzani, and P. Villa for useful
  suggestions and comments.}

\showacknow 



\bibliography{Bibliography-Avalanche+oscillations}

\newpage

\thispagestyle{empty}


\section*{Supplementary material}

# Supporting Information: Landau-Ginzburg theory of cortical dynamics: scale-free avalanches emerge at the edge of synchronization

**Serena di Santo**[a,b,c,1], **Pablo Villegas**[a,1], **Raffaella Burioni**[b,c], **and Miguel A. Muñoz**[a,2]

[a]Departamento de Electromagnetismo y Física de la Materia e Instituto Carlos I de Física Teórica y Computacional. Universidad de Granada. E-18071, Granada, Spain;
[b]Dipartimento di Fisica e Scienza della Terra, Università di Parma, via G.P. Usberti, 7/A - 43124, Parma, Italy; [c]INFN, Gruppo Collegato di Parma, via G.P. Usberti, 7/A - 43124, Parma, Italy

**SI1. Robustness of the results against changes in the dynamics.** In this appendix we confirm the robustness of the results and conclusions presented in the main part with respect to the modification of diverse ingredients and modeling details. In particular, we *first* discuss the full model including synaptic plasticity (as in the main text), but without truncating the equation for activity in a series expansion and, *second*, we consider inhibition as encapsulated in the well-known Wilson-Cowan equations (rather than synaptic plasticity) as a chief regulatory mechanism.

*Non-truncated excitatory-activity equation.* The dynamics in a single mesoscopic region of the cortex or "unit" is described by the full Wilson-Cowan equation (1) for the excitatory activity –such that the activity grows with the incoming current through a sigmoid response function– together with the Tsodyks-Markram TM model for synaptic plasticity (2):

$$\begin{cases} \dot\rho = & -a\rho + (1-\rho)\tanh(\rho R - \Theta) + h \\ \dot R = & \frac{1}{\tau_R}(\xi - R) - \frac{1}{\tau_D}\rho R. \end{cases} \quad [1]$$

In Figure 1, we illustrate that a linear-stability analysis reproduces a Hopf bifurcation scenario, as in the most relevant case (case A) discussed in the paper. When noise and spatial coupling are added, and the system is studied on a two-dimensional lattice, a synchronous irregular regime of network spikes, as well as an asynchronous irregular regime of nested oscillations –fully analogous to their corresponding counterparts in the the main text– are found, as graphically illustrated by the lower panels of Figure 1. This unveils the existence of a synchronization transition and confirms that the simplified truncated equation for the activity considered in the main text is a valid approximation of the full dynamics. Here we do not show a detailed analysis of the synchronization transition nor of the emergence of scale-free avalanches; but, let us remark that we have not found any substancial qualitative difference with respect to the case discussed in the paper in any of our exploratory checks.

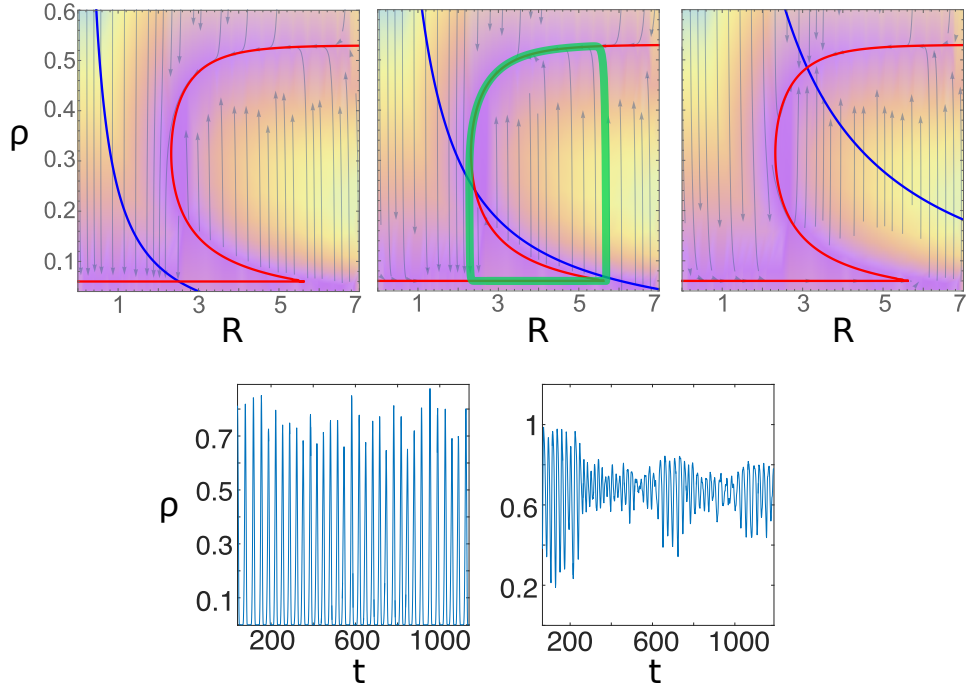

**Fig. 1.** Analysis of the model of Eq. 1. Upper panels: deterministic phase portrait with $\xi = 5, 12, 28$ (from left to right). respectively. showing a down state, a limit cycle and up state regimes, as in the case A of the main text. Other parameters are $a = 1, \tau_D^{-1} = 0.033, \tau_R = 500, \Theta = 0.34, h = 0.06$. Varying parameter values, it is possible to find either a similar Hopf bifurcation (case A) or a saddle node bifurcation (case B), as in the model with the truncated expansion. Lower panels: Temporal evolution of the total activity $\rho(t)$ on a two-dimensional lattice with $N = 64^2$ (after having introduced noise and coupling); in the (left) synchronous (network spiking) and in the (right) asynchronous (nested oscillations) regimes, respectively, revealing the presence of a synchronization phase transition in between the two regimes; parameter values: $\xi = 5$ and $\xi = 13$, respectively.

*Inhibition as main regulatory mechanism.* In this section we consider the full Wilson-Cowan equations (1), including both excitatory and inhibitory neural populations for each mesoscopic region or unit. In this case, inhibition plays the role of chief homeostatic mechanism, regulating the level of the overall network activity. More specifically, we consider a version of the Wilson-Cowan

dynamics, including also intrinsic noise as corresponds to large but finite (mesoscopic) regions. Such a model has been recently derived from an underlying microscopic model of spiking neurons in Ref. (3), and is described by the following set of stochastic equations for the densities of excitatory ($E$) and inhibitory ($I$) neurons:

$$\dot{E}_i = -\alpha E_i + (1 - E_i) \tanh\left[\omega_{EE} E_i - \omega_{IE} I_i + h\right] + \sigma \sqrt{\alpha E_i + (1 - E_i) \tanh\left[\omega_{EE} E_i - \omega_{IE} I_i + h\right]}$$
$$\dot{I}_i = -\alpha I_i + (1 - I_i) \tanh\left[\omega_{EI} E_i - \omega_{II} I_i + h\right] + \sigma \sqrt{\alpha I_i + (1 - I_i) \tanh\left[\omega_{EI} E_i - \omega_{II} I_i + h\right]}, \quad [2]$$

where $\alpha$ is the decay rate for the activity, $h$ is an external driving field, $\sigma$ is the noise amplitude, and $\omega_{ij}$ (with $i, j = E, I$) are the couplings between population $i$ and $j$ within a single unit; particularly important here is the auto-excitation coupling $\omega_{EE}$, which we take as a control parameter.

First of all, these equations are analyzed in the (noiseless) mean field limit. By increasing $\omega_{EE}$, the system exhibits a transition from a "down" state to an "up" state (see Fig. 2). Thus, a saddle-node bifurcation separates a state of high activity from a state of low activity, we found no track of a possible Hopf bifurcation. However, as soon as noise is switched on (i.e $\sigma \neq 0$), a noise-induced phenomenon appears: trajectories nearby the up-state fixed point, can escape from its basin of attraction as a result of fluctuations, and are then almost deterministically driven towards the down state, where a similar mechanism makes them escape with some probability. This phenomenon has been recently scrutinized in a remarkable work, where the role of non-normal forms in generating complex dynamics in general and avalanches in particular, has been emphasized (3).

This mechanism, generates in an effective way a noise-induced limit cycle between up and down states, which plays the same role as the deterministic limit cycle (Hopf bifurcation) of case A described in the main text. As a matter of fact, computer simulations of units described by Eq.(2), and coupled diffusively, give rise to the phenomenology illustrated in Fig.2: as the control parameter $\omega_{EE}$ is increased, the system undergoes a phase transition from a synchronous phase with very distinctive network spikes, to an asynchronous regime with nested oscillations, as it happens in the model with synaptic plasticity. Thus, also in this case, the phases are the same as in the main text and a synchronization transition appears between them.

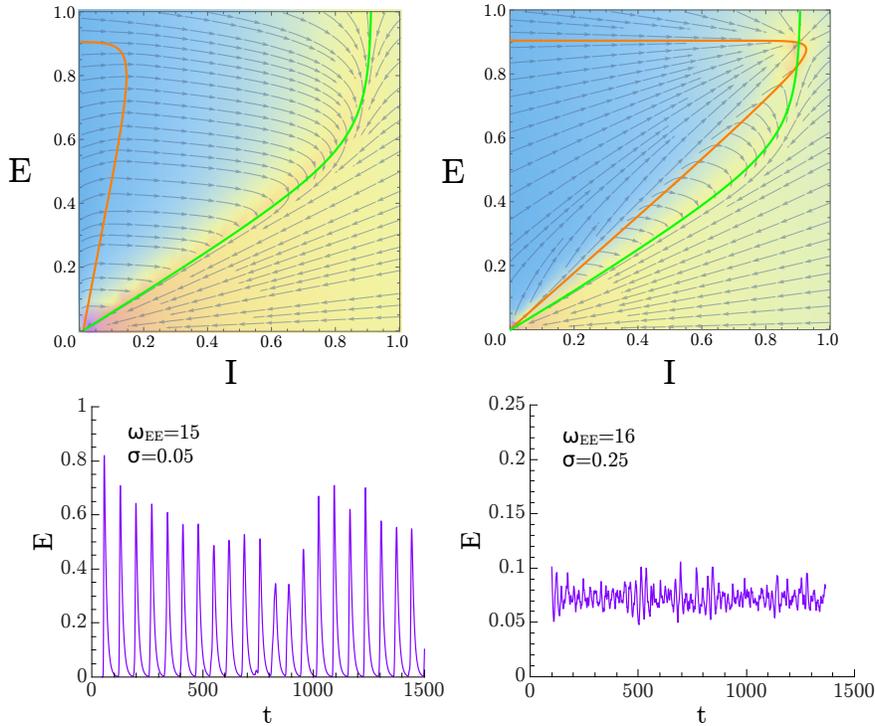

**Fig. 2.** Upper panels: mean-field analysis of the Wilson-Cowan set of Eqs. (2) describing both excitatory and inhibitory neural populations at each single unit, with parameters such that a noise-induced limit cycle (see (3)) in between a down and an up state can emerge once a non-vanishing noise is switched on. Observe that there is (left) a stable down-state fixed point ($\omega_{EE} = 4$) and a (right) stable up state ($\omega_{EE} = 16$); however the basin of attraction of the up state is small, and a relatively small fluctuation can induce the system state to go beyond the saddle-node line, where deterministic trajectories take the system toward the down state. In the lower panels we illustrate results of a computer simulation for a two-dimensional lattice of coupled noisy units, Eq.(2), corresponding to (left) synchronous/network-spiking and (right) asynchronous/nested-oscillation regimes. Parameter values: $D = 1, \omega_{EI} = 4.65, \omega_{IE} = 14.0, \omega_{II} = 2.8, h = 10^{-3}$ and $\alpha = 0.1$. Control parameter $\omega_{EE} = 15$ for SI regime and $\omega_{EE} = 16$.

**SI2. Robustness against changes in synaptic time scales, diffusion and noise.** As discussed in the main text, there are two possible scenarios according to the relation between the timescales for the recovery and depletion ($\tau_R$ and $\tau_D$, respectively). Namely, between the quiescent or 'down' state with $\rho^* \approx 0$ and the active or 'up' state with self sustained activity there exists a stable limit cycle (case A) or a regime of bistability (case B). Fixing parameter values while changing $\tau_R$ and $\tau_D$, it is possible to construct (mean-field or deterministic) a phase diagram showing the different possible cases that emerge when the control parameter $\xi$ is varied (cases A and B). As shown in Fig. 3 when the recovery time ($\tau_R$) is much bigger than the depletion time ($\tau_D$) the system is in the case A, while for bigger values of the depletion time ($\tau_D$) it falls into the case B with a transitions between up (active) and down (quiescent) states.

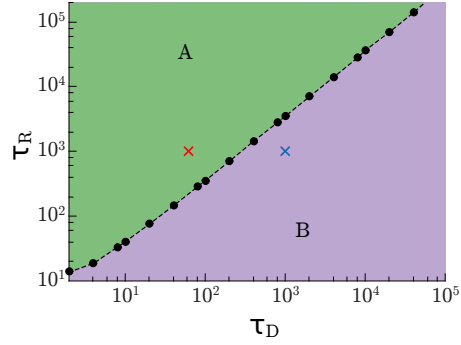

**Fig. 3.** Mean-field phase diagram showing the type of transition for different values of $\tau_R$ and $\tau_D$. Red (blue) cross show the particular case chosen in the Fig.1 of the main text for the case A (B). Parameter values are $a = 0.6, b = 1.3, h = 10^{-3}$.

We have also explored the behavior of the system against changes in the diffusion constant $D$. Figure 4 shows the phase diagram for different values of $D$ and some particular temporal series with the aim of characterize the different possible behaviors. As can be observed, there exists a transition from the synchronous irregular phase to the synchronous regular one for a wide range of $D$ values (.e.g. from $D = 0.01$ (red line) to $D = 2$ (violet line), and $D = 4$ (green line)). If $D$ is set to very large values, the system falls into the mean field expected behavior, switching from the network spiking regime to the up state. Similar conclusions are obtained, by fixing $D$ and decreasing the noise amplitude $\sigma$.

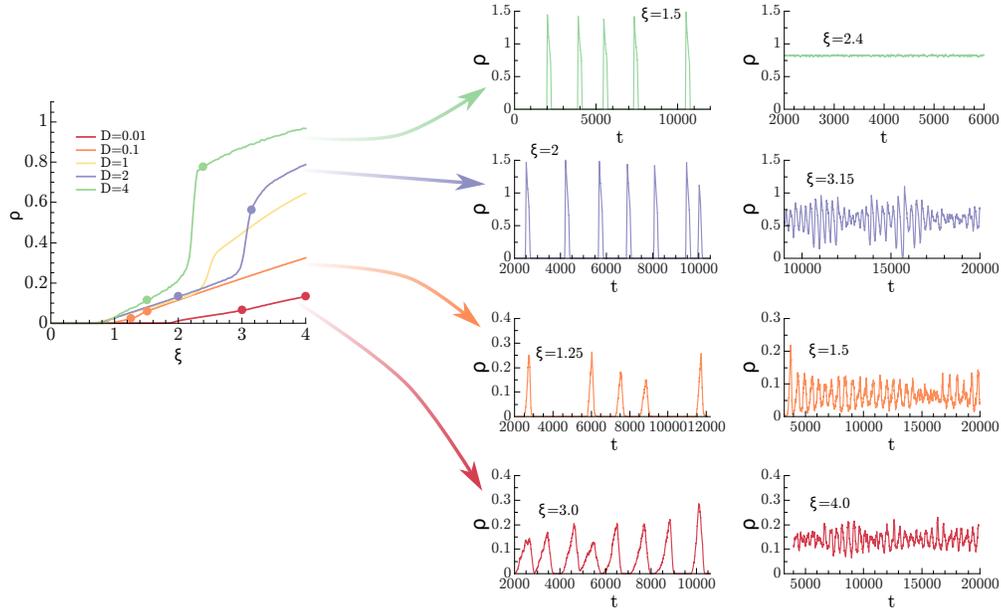

**Fig. 4.** Left panel: Order parameter as a function of the control parameter $\xi$ for various values of the diffusion constant $D$. Right panel: Temporal series for two particular values of $\xi$ for each value of $D$, (marked with colored points in the left panel), showing the expected behavior. Low values of $D$ show a transition between the synchronous irregular phase to the asynchronous irregular one, as in the main text (red, orange, violet and green line). Parameter values: $a = 1.0, b = 1.5, \tau_R = 10^3, \tau_D = 10^2, h = 10^{-7}$.

**SI3. The effect of long-range connections and network heterogeneity.** The detailed map of synaptic connections plays a central role in brain function (4). Even if most of the neuronal connections occur within the local neighborhood, long-range white-matter connectivity allows for information to be distributed and processed across the whole cortex. Such long-range connections comprise only about 10% of the total connections in the brain, but their role is crucial for brain functionality (4, 5).

*Small-world topology.* As the simplest possible approximation beyond a lattice of nearest neighbor connections, and consistently with (5), we built a small-world network, as done in the Watts-Strogatz model (6), by rewiring 10% of the links of a two-dimensional lattice. We explored the phase space of the model defined by Eqs. 1 and 2 of the main text, on this connectivity architecture (see Fig.(5) upper panel), and observed that the leading features described in the paper (i.e. phases and phase transitions) are preserved when long-range interactions are introduced. Indeed, as illustrated in the lower part of Fig.(5), our computational analyses reveal that the emergence of synchronous and an asynchronous phase, with a synchronization transition in between is a general intrinsic feature of our model, which is not modified by the small-world property of the network.

In any case, it is important to remark that even if the main phases remain unaffected, important details such as the extension of such phases, the specific shape of avalanches, the amplitude of nested oscillations, the broadness of the critical-like region etc. could be potentially sensitive to the introduction of network heterogeneity. Some of these aspects are explicitly illustrated in the forthcoming paragraphs.

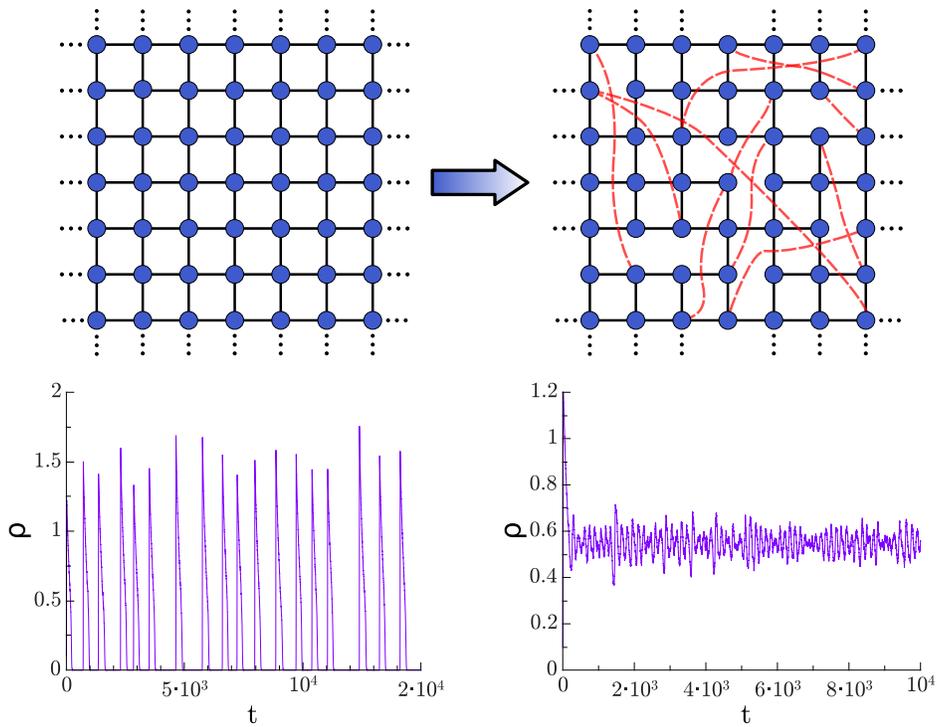

**Fig. 5.** Upper panel: sketch of the rewiring procedure defining a small-world network architecture: a ten percent of the links are rewired, in such a way that the average connectivity is preserved. Lower panel: Plots illustrating the results of computational analyses of the dynamics (main text, Eq. 1 and 2); in particular, temporal series of the global electrical activity in a small world lattice for two different values of control parameter $\xi$; (left, $\xi = 2.8$) synchronous/network-spike regime, and (right, $\xi = 2.96$) asynchronous/nested-oscillation regimes, respectively; a synchronization phase transition exists separating these two alternative regimes. Other parameter values: $a = 1.0, b = 1.5, \tau_D = 10^2, \tau_R = 10^3, h = 10^{-7}$.

*Clustered and heterogeneous networks.* Recent experimental analyses have scrutinized the effect of network heterogeneity in cultures of rat cortical neurons *in vitro* (7). In particular, Okujeni *et al.* where able to control the level of clustering by experimentally modifying the level of a given enzyme (protein Kinase C) that promotes neuronal aggregation. In this way, progressively more clustered networks were generated as the level of protein was increased (see Fig. 1 in (7)).

Keeping fixed other experimental conditions, Okujeni *et al.* found that in the case in which neurons are more homogeneously distributed in the substrate networks spikes appear much more sporadically than when the network is highly clustered (see Fig. 7, which is adapted from (7)), and that network spikes appear more clustered in time in this latter case. Thus, in conclusion, clustering promotes the generation of spontaneous network activity.

In order to model these experimental results, we developed a heterogeneous network in which we keep fixed the mean value of the parameter $a$ (that controls the decay of the activity at each single unit), but inducing some areas with low local values of $a_1$, i.e. with a smaller propensity for activity to decay (red nodes in Fig. 6), while in the rest of the network larger values of $a$, $a_2$, are considered (keeping the network-average value of $a$ constant).

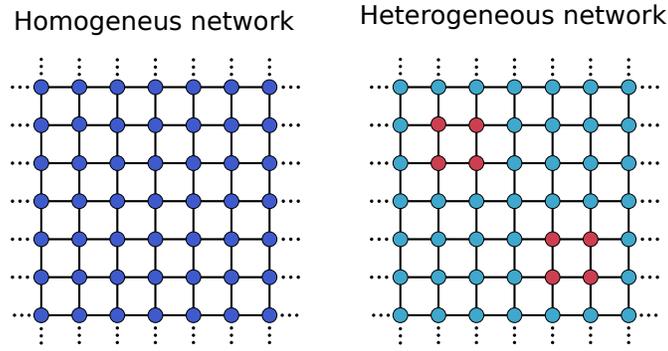

**Fig. 6.** Sketch of the considered networks: homogeneous to the left and heterogeneous/clustered to the right. In both cases the network-average value of the activity-decay parameter $a$ is taken to be equal. However, while in the homogeneous case the value of $a$ is constant across the network, in the heterogeneous one there are some areas (marked with red nodes) with a lower value of a.

As shown in Figure 7, the lower the local value of $a_1$, the more facilitated the emergence of spontaneous activity, leading the system closer and closer to the critical point or the asynchronous irregular phase, and reproducing quite remarkably the chief experimental observations of Okujeni *et al*.

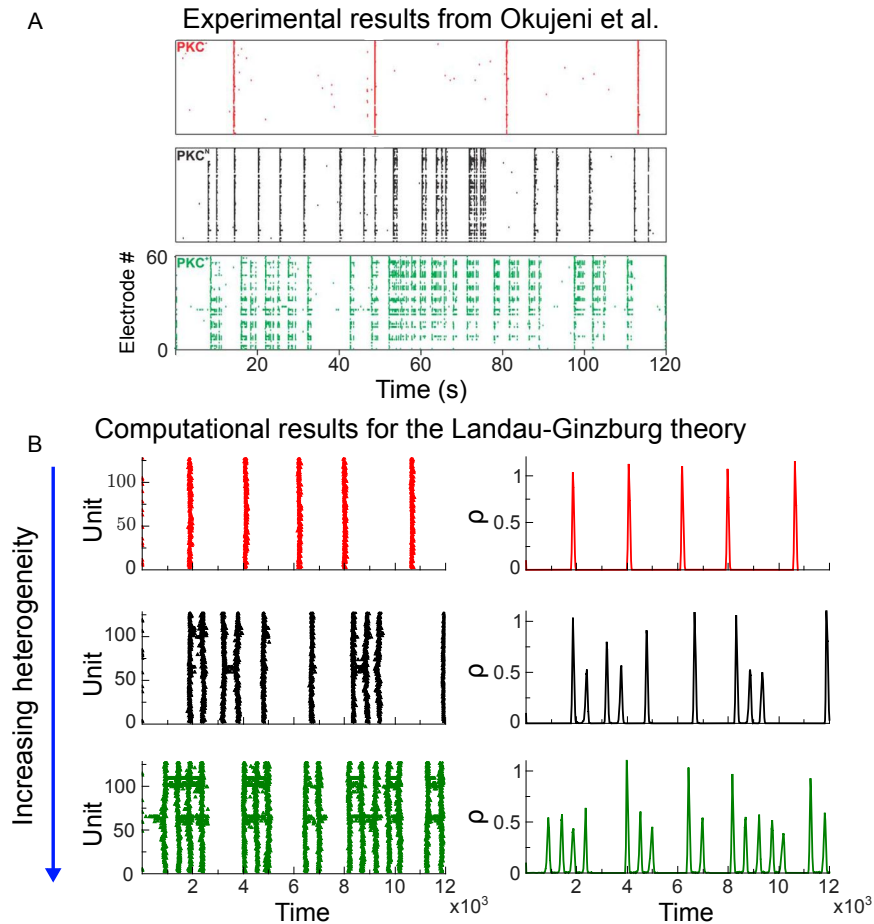

**Fig. 7.** Temporal series for different level of network clustering. Panel A shows the experimental results of Okujeni (7) (adapted figure from the original paper) for increasing levels of aggregation in a neural network. Panel B shows three temporal series for different levels of network clustering and a fixed value of $\xi = 1.2$. In the first one (red) the network is homogeneous with $a_1 = a_2 = 1$. Observe that smaller values of $a_1$ produce a more active network, in particular for $a_1 = -0.7$ (black) and $a_1 = -0.928$ (green). In both cases, the clustering facilitate the spontaneous activity. Other parameter values: $b = 1.5, \tau_D = 10^2, \tau_R = 10^3, h = 10^{-7}$.

Thus, in conclusion, our general model, equipped with an additional layer of network heterogeneity is able to reproduce specific empirical results.

**SI4. Detrended Fluctuation Analysis.** In this section we present an additional type of analyses to discriminate whether the system lays at a critical point or in either the subcritical or the supercritical phases. The method is based on the fact that, at the critical point of a continuous phase transition, the (time-dependent) order parameter, as measured in any finite system, shows long-range temporal correlations (i.e. long-memory effects), which can be quantified by measuring its Hurst exponent (8). The Hurst exponent of a time series is a measure of the dispersion of a process on a scaling support. For example the Hurst exponent of an uncorrelated signal (white noise) is $\alpha = 1/2$, since the root mean square translation distance after $n$ steps of a Wiener process, i.e. an unbiased random walk (the process obtained by integrating white noise), is proportional to $\sqrt{n}$. For correlated signals (colored noises) one expects bigger Hurst exponents (as a reference, $\alpha \simeq 1$ is found for pink noise). The Hurst exponent can be calculated by splitting the time series into adjacent windows, plotting the square-root displacement from the mean as a function of the window size and evaluating the exponent of the resulting power law (see below). More specifically, "detrended fluctuation analysis" (DFA) is a technique for measuring the Hurst exponent in non-stationary time series: the "detrending" operation allows to remove fictitious memory effects related to non-stationarity, and the method basically consists in subtracting the local "trend" (usually using a linear fit approximation) of the signal before performing the analysis on each window (9, 10). DFA consists of two steps: the data series $\rho(t)$ is shifted by its mean $\bar{\rho}$, and integrated (cumulatively summed) in time:

$$\mathcal{P}(\tau) = \sum_{t=1}^{\tau} \left(\rho(t) - \bar{\rho}\right);  \quad [3]$$

then segmented into $k$ windows of various sizes $n$, and for each window size, a fluctuation function $F(n)$ is calculated, as

$$F(n) = \sqrt{\frac{1}{T} \sum_{h=1}^{k} \sum_{\tau=1}^{n} \left(\mathcal{P}^{(n)}(\tau + (h-1)n) - X_{\mathcal{P}}^{(n)}\right)^2}  \quad [4]$$

where $X_{\mathcal{P}}^{(n,h)}$ is the linear regression of $\mathcal{P}^{(n)}(\tau)$, with $\tau \in [(h-1)n, hn]$, the superscript indicates the dependence on the window size $n$ and $T = kn$ is the total length of the time series. If $F(n) \sim n^{-\alpha}$, then $\alpha$ is the Hurst exponent (9, 11).

We performed a DFA on the global signal $\rho(t)$ coming out of our computer simulations for different values of the control parameter $\xi$ (in the synchronous and asynchronous phases as well as at the critical point). Results are shown in Fig.8: (i) fluctuations in the asynchronous phase grow approximately as the square root of the window length, as expected for (uncorrelated) white noise; (ii) in the synchronous phase, above a certain characteristic length, the dependence is very weak, remarking the existence of a certain degree of order, i.e. a characteristic time scale at which there is order, i.e. synchronization; (iii) just at the critical point the growth of the fluctuations is anomalously large, confirming the existence of long-range correlations: a signature of criticality. Therefore, from the global activity signal we are able –through a DFA analysis– to conclude that long-range correlations, characteristic of criticality, emerge at the transition point.

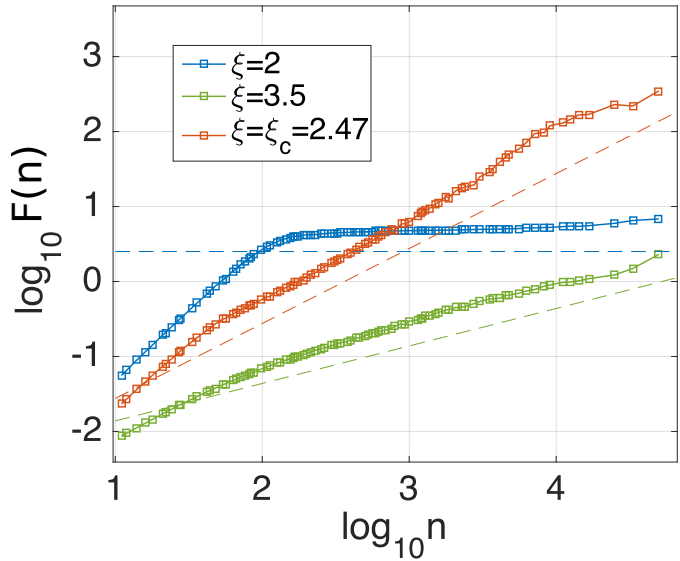

**Fig. 8.** Detrended fluctuation analysis of the macroscopic signal for different values of the control parameter $\xi = 2$ (synchronous phase) $\xi = 2.47$ (critical point), and $3.5$ (asynchronous phase), respectively. Close to the transition point the DFA shows an Hurst exponent close to 1 (orange dashed line) implying long-range autocorrelations, a fingerprint of criticality, while the white noise value $\alpha = 1/2$ (green dashed line) emerges in the asynchronous regime, and an asymptotically almost flat curve is obtained in the synchronous phase, revealing the existence of a characteristic time scale. Parameter values are taken as in the main text and $N = 2^{14}$.

**SI5. The nature of nested oscillations.** In order to unveil the nature of the nested oscillation (asynchronous irregular, (AI)) phase and to determine whether it is a finite size effect or it survives in the thermodynamic limit, the existence of a second phase transition separating it from the up state is investigated here. In other words: are the asynchronous irregular phase and the up-state phase two different phases, or are they just the same phase, with only a quantitative difference in the amplitude of the variability around the mean value? As we illustrate in what follows the correct answer is this second one.

In principle, these two regimes show a qualitative difference: in the AI phase each single unit keeps switching between the *on/up* and *off* states and there exists a macroscopic fraction of *off/down* sites, whereas in the active phase units are permanently in the *on* state and, even if fluctuations might lead some unit to the *off* state, the macroscopic fraction of them vanishes (see Supplementary Movie). The fraction of inactive units, $\rho_0$, can thus be chosen as an order parameter for the putative phase transition between the AI and the active phase. In Figure 9 we plot the average over time of $\rho_0$ in function of $\xi$ and we verify that this alternative order parameter detects the same phase transition already characterized in the main text, by employing synchronization order parameters. This implies that, in the large system-size limit, there exists no macroscopic difference between the asynchronous/nested-oscillation regime and the up state. Therefore, the nested oscillations can be understood as the result of partial synchronization of local regions; the superposition of a few regions gives rise to complex waves as those in Fig.2 (A3) of the main text. However, when the system becomes progressively large, the number of such locally synchronized regions grows, and their interference leads to a standard up state, in which fluctuations around the mean density decay as a function of the system size.

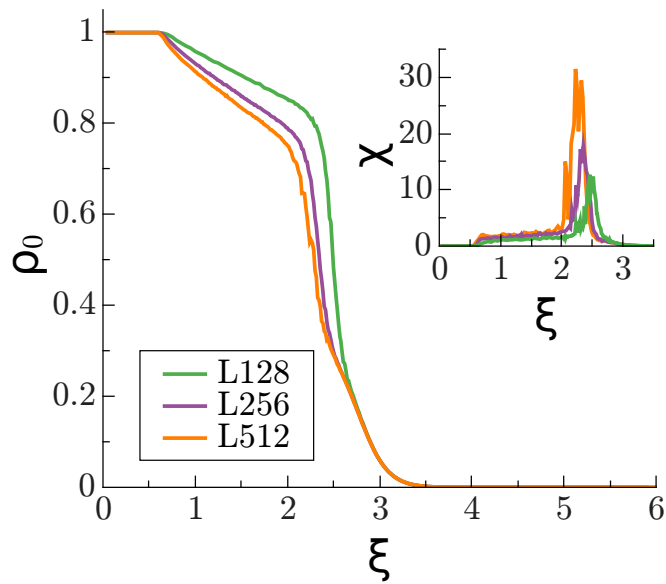

**Fig. 9.** Main: averaged fraction of inactive sites in the system $\rho_0$ as a function of the control parameter $\xi$, revealing the presence of a phase transition. Inset: Variance over runs of a given fixed duration of the average value of the control parameter multiplied by $\sqrt{N}$ in order to highlight possible deviations with respect to central limit theorem (CLT); as a result of which, a decay with $\sqrt{N}$ is expected; thus multiplying by $\sqrt{N}$ a convergence to a constant should be expected if the CLT holds. Observe, however, the increase in peak height as the system size is enlarged revealing a violation of the CLT, as expected at the critical point of a second order phase transition. Note that for all the system sizes the peaks are located approximately in the same spots as in Fig.4 in the main text; thus $\rho_0$ is an alternative order parameter that leads to the same results as the previously considered synchronization order parameters: it detects the synchronization phase transition, and reveals that there is no difference between the AI and the active phase in the limit of infinitely-large network sizes.

**SI6. Definition and analyses of avalanches.**

***On the definition of spiking events.*** In the analysis of avalanches presented in the main text, a particular (and reasonable) criterion has been chosen to convert each local (continuous) time signal into a discrete series of spikes, allowing to build up the raster plot, from which avalanches are measured using the standard experimental protocol consisting on a time-clustering the events (12).

Here, we employ a different criterion to define spikes, thereby illustrating the robustness of our main findings against such a choice. In particular, the alternative discretization criterion is sketched in Figure 10 and is as follows: a threshold $\theta$ is established at each single unit, and every unit is declared to be in its *spiking* (or "on") state whenever its activity is over threshold. Thus, the main difference with the protocol in the main text is that now, in between two-consecutive time steps in which the unit is below threshold, the site is considered to be "on" not just at one time step (at its maximum of activity, as in the method of the main text), but possibly during many time steps, in a full time interval.

Considering these spiking events, avalanches are defined through the same experimentally inspired protocol that we used in the main text; the size of an avalanche is simply the number of spike counts during an avalanche. Figure 10 shows that the avalanche size distribution at the critical point is preserved by employing this alternative definition of the spikes. Moreover if a (random) subsampling of the units is performed, the distribution keeps following a power law which is roughly consistent with the experimentally measured exponent $3/2$.

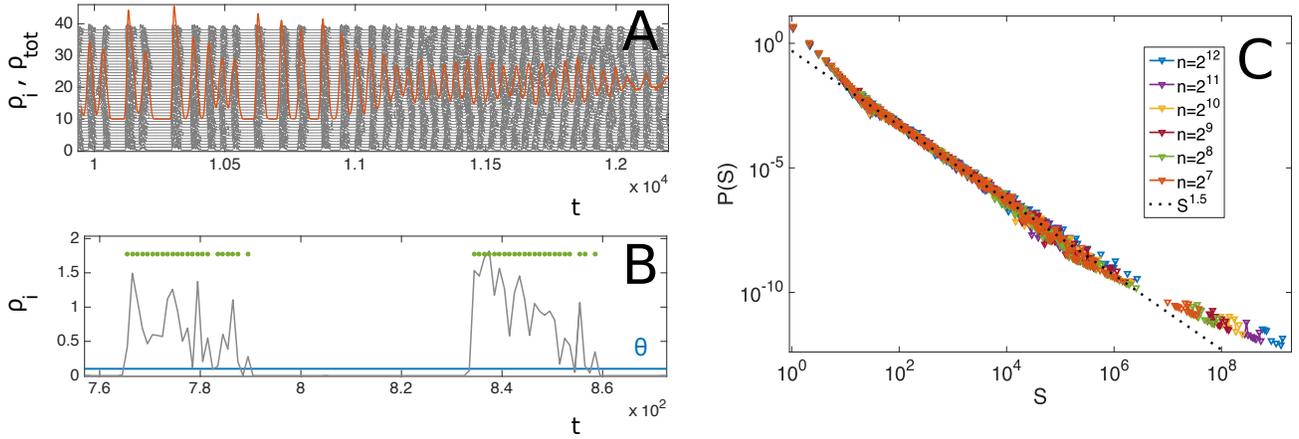

**Fig. 10.** A. Sample of local temporal signals are plotted in grey (shifted for convenience of visualization) together with the global signal $\rho(t)$ (shifted and rescaled) represented in red color, for a lattice of $N = 64^2$ units. B. illustration of the alternative method employed here to define "spike events" from a local temporal signal. Green dots represent times during which the unit is in its spiking state; obviously, a discrete (integration) time is required to have a finite number of spikes per interval of local activity. C. Distribution of avalanche sizes for various subsampling trials, using the criterion sketched in B to define events. The black dotted line is the power law with exponent $3/2$ plotted as a guide to the eye.

***Threshold effects.*** The standard empirical method to detect avalanches, as defined in (12), is intrinsically affected by some arbitrariness in parameter choices, that has been already discussed in the literature (12, 13). In particular, one arbitrary parameter is the threshold value $\theta$ above which the state "on" or "spiking" is declared [*]. The common belief is that if, as a matter of fact, the system is scale-invariant, this value should not affect the large scale properties, such as exponent values. However, one has to be particularly careful with any thresholding procedure. For example, while for a standard one-dimensional random-walk process the avalanche exponents are independent of the threshold value chosen, this is not the case for other stochastic processes, e.g. birth-death processes (14). Moreover the threshold value should not be chosen too high relatively to the amplitude of the signal in order to avoid splitting an event into multiple (correlated) ones (15).

Recording a spike every time that the system crosses a very small threshold exposes the measurements to the effects of small fluctuations around the origin, induced by the multiplicative demographic noise term in Eq.1 in the main text (see Fig. 11). In order to avoid such a problem, it is possible to set also a second threshold value $A_{min}$ for the minimal area for a spiking event to be considered as such; below such a threshold, activity is considered just a noise effect and, hence, disregarded. As illustrated in Figure 12 the statistics of avalanches does not depend significantly on the value chosen for such a threshold, $A_{min}$.

---

[*]In order to avoid spurious effects and consistently with the definition of avalanches as activity propagating marginally before falling into an absorbing state ($\rho = 0$), we choose a small value $\theta \ll 1$ (namely $\theta = 10^{-4}$).

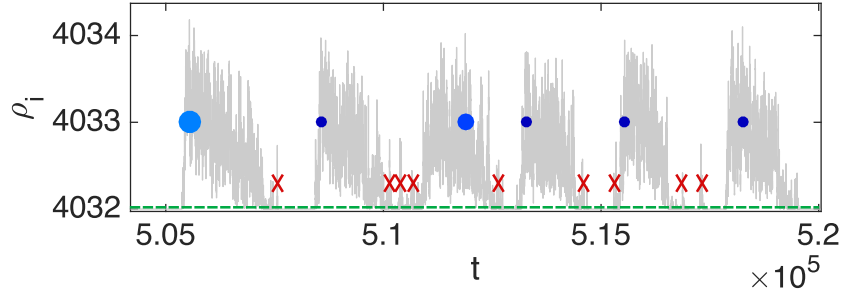

**Fig. 11.** Illustration of small events in the signal of the activity of one unit. Events covering a very small area (marked with red crosses) are neglected, while proper spikes are marked with blue full dots (smaller darker dots correspond to spikes with smaller area).

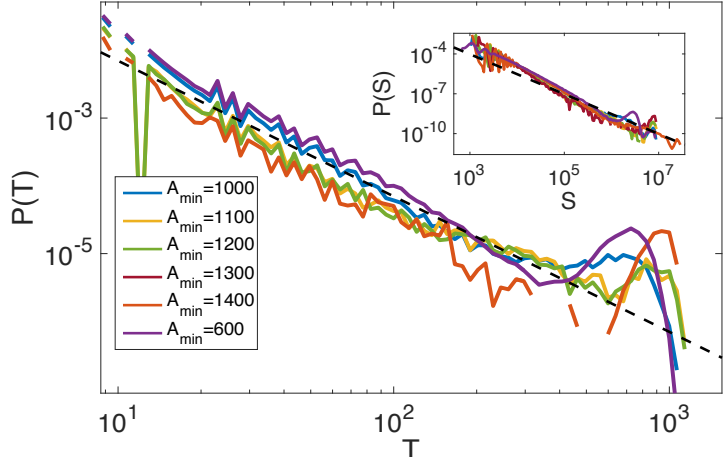

**Fig. 12.** Independence on the avalanche exponent values on the value chosen for the threshold on the minimal area $A_{min}$, used to declare "activity". Apart from the details of the substructure of the distribution, no qualitative dependence can be found by varying the value of $A_{min}$, both in avalanche-time (main plot) and avalanche-size (inset) distributions.

***Experimentally inspired procedure to measure avalanches.*** Here, we briefly explain the steps necessary to perform a measure of an avalanche's size and duration, starting from a raster plot, as usually done in experimental setups (12). First of all, the Inter-Spike-Interval ($ISI$) is measured, as the average time interval between two consecutive spikes of (whichever 2 elements of) the network. Then the raster plot is divided in contiguous bins of width equal to the $IEI$ (see Fig.5C in the main text). A bin is considered "empty" if no events are reported within it, and "occupied" otherwise. Consecutive series of occupied bins, preceded and ended by an empty one, define an avalanche. The avalanche duration is just the time interval between the preceding and the ending empty bins, and avalanche size is the total number of spikes that occurred in that time-interval. Since the individual signal in our analysis stems from a coarse grained section of neural tissue, we assign a weight to each event, representing the number of spikes within it, and determined by the integral of the signal during the event (see Fig.5A). Thus the only difference between the procedure we employ and the experimental one (12) is that the size of an avalanche is defined in our case as the *weighted* sum of the events during an avalanche.

***System-size dependence.*** At the critical point of a phase transition, scale-invariant behavior is expected to be only limited by system size. This effect, which has been reported to be observed in experiments on neural avalanches (12), is also a hint in favor of true scale invariance, since finite size scaling holds when the system is at its critical point. In Figure 13 avalanche size and duration distributions are compared for various system sizes; as expected, larger systems show larger avalanches with progressively larger cut-off scales, while the overall size (resp. time) distribution keeps following a power law trend with the usual exponents $\tau = 3/2$ (resp. $\alpha = 2$; see inset).

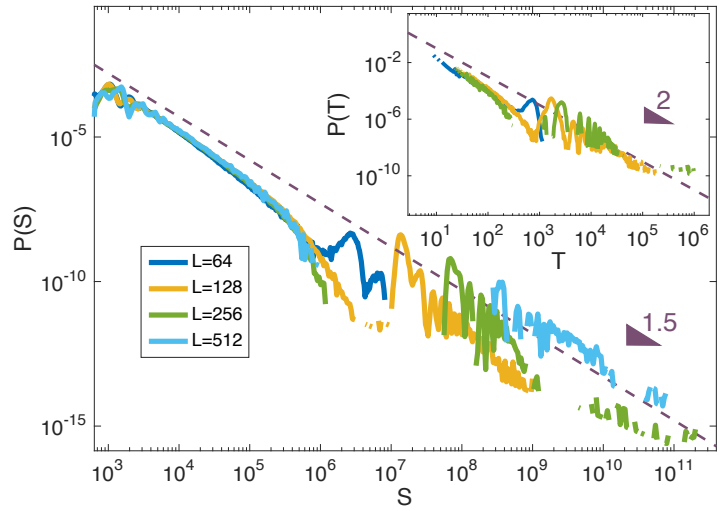

**Fig. 13.** Avalanche size $S$ and duration $T$ distributions for system sizes $N = 2^{12}, 2^{14}, 2^{16}, 2^{18}$. Although finite size scaling is not perfect, as the system size grows progressively larger, larger avalanches are found. Dashed lines represent an hypothetic power law trend with $\tau = 3/2$ and $\alpha = 2$.

**SI7. Oscillations coexisting with scale invariance.** Usually, scale-free avalanches of activity can be measured at the critical point of an *absorbing-state phase transition*. When the concept of "avalanche" is employed to describe the critical point of a synchronization phase transition, the marginal oscillatory nature of the system introduces a characteristic time scale in the synchronous phase –i.e. the period of the oscillations– which, in principle, is in contrast with the idea of scale-invariance. However, the two concepts can coexist –at least within certain limited scales– as illustrated in Figure 14. It shows that the structure (e.g. the peaks) in the avalanche-time distribution (inset) corresponds to the period of oscillation of a macroscopic variable (the total number of spikes, in the main plot); for instance, an isolated network synchronization event has a typical duration of 2000 (in arbitrary units), a sequence of two, about 5000, etc. On the other hand, the whole distribution, once these peaks are ignored can be approximately described as a power law with the expected exponent value.

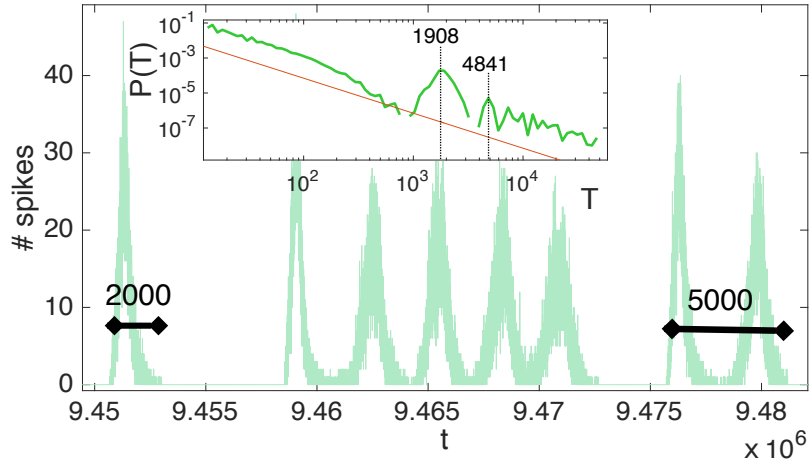

**Fig. 14.** Analysis of the structure underlying the avalanche-duration distributions. The main figure shows the total number of spikes at time t. Irregular oscillations of the global activity can be recognized, as the system is close to the edge of the synchronization phase transition. The characteristic period of an isolated oscillation corresponds to the peak in the avalanche duration distribution, while its multiples correspond to smaller peaks. System size $N = 128^2$.



\end{document}